\documentclass[a4paper,fleqn,usenatbib]{mnras}

\usepackage[T1]{fontenc}
\usepackage{ae,aecompl}

\usepackage{graphicx}	% Including figure files
\usepackage{amsmath}	% Advanced maths commands
\usepackage{amssymb}	% Extra maths symbols
\usepackage{subcaption}
\usepackage{afterpage}  % For dealing with frustrating floats
\usepackage{txfonts}

\graphicspath{{./figs/}}
\DeclareGraphicsExtensions{.pdf}

\newcommand{\aref}[1]{\hyperref[#1]{Appendix~\ref{#1}}}

\title[Observability of intermittent radio sources]{Observability of intermittent radio sources in galaxy groups and clusters}

\author[Patrick M. Yates et al.]{
Patrick M. Yates,$^{1}$\thanks{E-mail: patrick.yates@utas.edu.au}
Stanislav S. Shabala,$^{1}$
Martin G. H. Krause$^{2,1}$
\\
$^{1}$ School of Natural Sciences, Private Bag 37, University of Tasmania, Hobart, TAS 7001, Australia\\
$^{2}$ Centre for Astrophysics Research, University of Hertfordshire, College Lane, Hatfield, Herts AL10 9AB, UK \\
}

\date{Accepted XXX. Received YYY; in original form ZZZ}

\pubyear{2018}

\begin{document}
\label{firstpage}
\pagerange{\pageref{firstpage}--\pageref{lastpage}}
\maketitle

\begin{abstract}

    We have carried out numerical hydrodynamic simulations of radio jets from active galactic nuclei using the PLUTO simulation code, with the aim of investigating the effect of different environments and intermittency of energy injection on the resulting dynamics and observable properties of the jet-inflated lobes.
    Initially conical jets are simulated in poor group and cluster environments.
    We show that the environment into which a radio jet is propagating plays a large role in the resulting morphology, dynamics and observable properties of the radio source.
    The same jet collimates much later in a poor group compared to a cluster, which leads to pronounced differences in radio morphology.
    The intermittency of the jet also affects the observable properties of the radio source, and multiple hotspots are present for multiple outburst jets in the cluster environment.
    We quantify the detectability of active and quiescent phases, and find this to be strongly environment-dependent.
    We conclude that the dynamics and observational properties of jets depend strongly on the details of energy injection and environment.
\end{abstract}

\begin{keywords}
hydrodynamics -- galaxies: active -- galaxies: jets -- radio continuum: galaxies
\end{keywords}

%%%%%%%%%%%%%%%%%%%%%%%%%%%%%%%%%%%%%%%%%%%%%%%%%%

%%%%%%%%%%%%%%%%% BODY OF PAPER %%%%%%%%%%%%%%%%%%

\section{Introduction}

It is well accepted that outflows from active galactic nuclei play an important role in slowing cooling flows \citep[see reviews by e.g.][]{McNamara2007, Alexander2012, Fabian2012}.
The details of how the jet energy couples with the environment is still an open problem, and a detailed prescription is needed for semi-analytic galaxy formation models \citep[e.g.][]{Croton2006,Shabala2009,Raouf2017} and cosmological galaxy formation simulations \citep[e.g.][]{Vogelsberger2014,Schaye2015,Kaviraj2016}.
Intermittent jet activity is required in order to maintain the heating/cooling balance of active galactic nuclei and their host galaxies \citep{Heckman2014}, and is supported through observational evidence of double-double radio galaxies \citep[e.g.,][]{Schoenmakers2000a}.

The first basic morphology models of FR II \citep{Fanaroff1974} radio sources were introduced by \citet{Scheuer1974} and \citet{Blandford1974}, which both proposed a relativistic outflow from a central region.
\citet{Begelman1989} proposed that the cocoons surrounding this relativistic outflow were overpressured with respect to the intergalactic medium, and \citet{Falle1991} showed that these outflows have self-similar expansion.
An analytic self-similar expansion model to produce the complete FR II morphology was developed by \citet[][the KA model]{Kaiser1997}, which has radio sources expanding into an environment with a smooth density profile given by a power-law.
This was extended by \citet[][the KDA model]{Kaiser1997a} to include energy losses from synchrotron processes due to relativistic electrons in the jet cocoon, allowing the calculation of radio emission from the radio source.
An alternative model for the spectral evolution of radio sources was proposed by \citet{Manolakou2002}, which calculates the first-order Fermi acceleration of the electrons at the termination shock as opposed to assuming an electron distribution for the cocoon.
The self-similarity assumption used in the KA model does not hold for small ($<1\,\mathrm{kpc}$) scales, and was extended by \citet{Alexander2006a} to better model the uncollimated to collimated transition.
The extended model introduces the length-scale $L_1$, which relates the jet density and environment density, and these characteristic length-scales are expanded by \citet[][see Sect 2]{Krause2012}.
Recently, semi-analytical models for the evolution of radio sources have been developed \citep{Turner2015,Hardcastle2018} which relax the self-similarity assumptions, and allow arbitrary environments to be specified.

Different types of radio sources are found in different environments \citep{Longair1979}: while more powerful, edge-brightened FR II radio sources tend to reside in lower mass halos, the less powerful FR Is (edge-darkened) are more frequent in massive galaxy clusters.
There is not a one-to-one mapping, but there is a tendency for FR I hosts to have lower accretion onto their supermassive black holes and less star formation than FR IIs \citep{Buttiglione2009,Hardcastle2013b}, consistent with a dependence on the mass of the dark matter halo.

Apart from the basic Fanaroff-Riley morphological classification, one expects the radio morphology and luminosity to be strongly affected by the environment: an FR II jet in a rich cluster will quickly come into sideways pressure equilibrium \citep[see][]{Hardcastle2013}, but the cluster atmosphere will still collimate the jet relatively early, and thus produce a narrow beam.
The high cluster density will ensure that the lobes will be bright radio emitters.

The gas pressure in a poor group is much lower.
Correspondingly, the length scale $L_2$ \citep[see][]{Krause2012}, where the lobes come into pressure equilibrium with the environment is much larger.
The lobe pressure can therefore be lower than in the cluster case, while still overpressured with respect to the ambient gas.
The latter makes the lobe dynamics different from the cluster case; in groups, jets will be collimated later and hence wider.

It is clear from these considerations that it should be much more difficult to observe a jet in a poor group.
Yet, the AGN feedback might be crucial just for these halos, as the more massive "green valley" galaxies, that transition from a state of high star formation rate to quiescence, are usually found in such haloes \citep[e.g.][]{Alatalo2014,Alatalo2016}.
The advent of a new generation of observing facilities means that their study might become feasible in the near future.

The environmental dependence of the radio properties of ambient pressure-collimated jets has been investigated by \citet[][HK13]{Hardcastle2013}.
For parameters representing the range from poor groups to rich clusters, it was found that the range in radio luminosity for FR II radio sources that have been evolved for about $10^8$ years spans roughly one order of magnitude for a given jet power.

To flesh out the difference in observability, we concentrate here on two environments at the extremes of the parameter range: one with a dark-matter halo mass of $3 \times 10^{12}\,\mathrm{M_\odot}$, and one that is a hundred times more massive, with isothermal IGM/ICM gas in hydrostatic equilibrium with the dark matter halo.
We inject jets with FR II parameters, a single jet power and observationally motivated duty cycles, and calculate the luminosity evolution as well as the surface brightness to judge the observability of the simulated sources.

\section{Simulation setup}

The simulations presented in this paper are carried out using version 4.2 of the PLUTO\footnote{http://plutocode.ph.unito.it/} code for computational astrophysics \citep{Mignone2007}.
PLUTO has been used for other AGN jet simulations \citep{Hardcastle2013,Hardcastle2014,Nawaz2014,Mukherjee2016}, and is adept at handling high-velocity astrophysical flows in the presence of discontinuities.
Using PLUTO we evolve the equations of hydrodynamics in 2D axisymmetry with the ``hllc'' Riemann solver, linear reconstruction, the ``minmod'' flux limiter, and second-order Runge-Kutta time-stepping.

\subsection{Non-dimensionalisation}\label{sec:non-dim}

The simulations are non-dimensionalised using length scales corresponding to morphological changes in an initially conical jet \citep{Alexander2006a,Krause2012}.
The unit length scale is taken to be $L_1$, 
\begin{equation}
    L_1 = 2 \sqrt{2} \left( \frac{Q}{\rho_\mathrm{x} v_\mathrm{jet}^3} \right) ^ {1/2}.
\end{equation}
Here $\rho_\mathrm{x}$ is the ambient density (for a constant density environment), $Q$ is the jet kinetic power and $v_\mathrm{jet}$ is the jet velocity.

Three additional length scales related to $L_1$ are presented in \citet{Krause2012}, $L_{1a}$, $L_{1b}$, and $L_{1c}$, corresponding to the jet recollimation scale; cocoon formation scale; and termination shock point for an uncollimated jet, respectively.
$L_{1b}$ is the scale in a constant density environment where the density of an initially overdense, conical jet drops below the ambient density.
We choose the unit density in simulations to be the ambient density at distance $L_{1b}$ from the core, where
\begin{equation}
    \frac{L_{1b}}{L_1} = \left( \frac{1}{4 \Omega} \right)^{1/2}.
\end{equation}
Here $\Omega = 2 \pi (1 - \cos{\theta_\mathrm{jet}})$ is the solid angle of a conical jet with half-opening angle $\theta_\mathrm{jet}$.

With the addition of $c_\mathrm{x}$, the external sound speed, as the unit speed, all remaining unit quantities such as time and energy can be calculated.
The scaling of these parameters for the two environments we probe is included in \autoref{tbl:simulation-runs}.

\subsection{Simulation grid and resolution}

A two-dimensional spherical polar grid is used for the simulations presented in this paper, similar to that used by HK13.
\citet{Hardcastle2014} showed that general large-scale lobe dynamics from three dimensional simulations are reproduced in two dimensional ones, justifying our choice of dimensionality.
By trading an extra spatial dimension for increased resolution in the remaining two, our simulations are able to resolve the self-collimation of the jets with a realistic computational complexity requirement.
Our choice of dimensionality and the impact this has on the simulations are discussed in \autoref{sec:disc-dimensionality}.

Six grid patches each containing a number of equally spaced grid cells in $r, \theta$ allow the simulation to properly capture the physics of the jet while reducing the computational power required by having coarser resolution in areas of the simulation domain where changes to the hydrodynamical quantities ($\rho, P, \vec{v}$ etc.) are slow.

The two radial grid patches are 64 grid cells from $r=1.0$ to $r=2.0$ (simulation units), and 2000 grid cells from $r > 2.0$ to the end of the simulation domain.
The inner radial grid patch provides the high resolution necessary to accurately capture the injection of the jet onto the simulation grid.

The three azimuthal grid patches are 64 grid cells from $\theta = 0\,-\,5$ deg, 128 grid cells from $\theta = 5\,-\,17$ deg, and 256 grid cells from $\theta = 17\,-\,90$.
The inner azimuthal grid patch provides a consistently high resolution across the jet head, with approximately 15 grid cells across a $2\,\mathrm{kpc}$ wide jet head at $100\,\mathrm{kpc}$ from the core.
The second azimuthal grid patch extends high resolution out to $\theta = 17^\circ$.
Our jets are injected along the $\theta = 0$ plane with a half-opening angle of $\theta_\mathrm{jet} = 15^\circ$, and simulation quantities are slowly varying within the third azimuthal grid patch.
Hence our resolution at the largest angles is sufficient.

High resolution single outburst simulations are performed in the poor group and cluster isothermal NFW environments to verify that the jet dynamics are captured correctly.
These simulations have 12000 cells in the outer radial grid patch, and 200, 250, 250 cells in the first, second and third azimuthal grid patches respectively.
The evolutionary tracks of these simulations are shown in \autoref{fig:pd-c-extra} and \autoref{fig:pd-pg-extra}, and they are discussed in \autoref{sec:discussion}.

Reflective boundary conditions are used for the lower and upper radial boundaries, which are needed in order to conserve mass in the simulation.
An axisymmetric boundary is used along the $\theta = 0$ axis, while an equatorially symmetric boundary is used along the $\theta = \pi / 2$ axis.
This equatorially symmetric boundary approximates the presence of a counter-jet.
Simulating the full plane has the advantage of removing azimuthal boundary conditions, as shown by HK13, however here we find very little turbulence near the $\theta = \pi / 2$ boundary, and therefore no significant pressure fluctuations across that boundary.
Hence our approach is robust.

\subsection{Jet injection}\label{sec:jet-injection}

The jet is injected as a conical mass inflow boundary condition on the lower radial boundary where $0 \le \theta \le \theta_\mathrm{jet}$.
A half-opening angle of $15^\circ$ is chosen because it produces FR II morphologies in constant density environments \citep{Krause2012}.
We inject a pressure matched jet, corresponding to cells on the injection boundary at $r=1$ in simulation units having a density $\rho=\rho_\mathrm{jet}$, $p=p_\mathrm{x}$, and a radial velocity $v_\mathrm{r}=M_\mathrm{x} c_\mathrm{x}$ for $\theta \le \theta_\mathrm{jet}$, while cells with $\theta \ge \theta_\mathrm{jet}$ have a reflective boundary condition imposed.
A scalar tracer field is injected with a value of $1.0$ in the jet and $0.0$ elsewhere.

The physical injection radius is the length scale $L_1$ for the environment, which ranges from $0.36$ to $17.8\,\mathrm{kpc}$ for the simulations presented in this paper.
Injecting the jet at these distances from the core is sufficient for exploring the interaction of the jet with the homogeneous intracluster medium.

In this work, we choose the jet kinetic power to be $Q = {10}^{37}\,\mathrm{W}$, typical of low-power FR IIs \citep{Turner2015}.
The FR I/II separation is not very sharp in jet power.
The power we use here is in the transition region, so that both sources in clusters, which tend to have lower power, and groups, which may frequently have higher jet power can be addressed.

The computation time required for the simulations increases with external Mach number as $M_\mathrm{x}^{3/2}$ due to the use of $L_1$ as the unit length.
Following the work of \citet{Hardcastle2013}, we adopt an external Mach number of $M_\mathrm{x}=25$, which provides a suitable trade-off between realistic jet dynamics and computation time.
We validate our approach by also performing higher Mach number simulations (75 and 200) for each isothermal NFW environment, with the same simulation grid as the high resolution Mach 25 simulations; these are discussed in \autoref{sec:discussion}.

\subsection{Environment}\label{sec:env}

A key feature of our simulations involves the environment into which the jet is propagating.
We wish to study jets at redshift $z\sim0$, and adopt the Planck15 cosmology (Planck Collaboration \citeyear{Planck2016}) with $H_0 = 67.7\,\mathrm{km}\,\mathrm{s}^{-1}\,\mathrm{Mpc}^{-1}$, $\Omega_\textrm{M}=0.307$.
In this work, we use two types of gas distributions: isothermal gas in hydrostatic equilibrium with the dark matter potential (the isothermal NFW profile); and an isothermal beta profile \citep{King1962} which reasonably describes observations of low-redshift clusters.
The isothermal NFW gas density profile is the main focus for this paper, and the (similar) isothermal beta profile results are discussed in \autoref{sec:discussion}.

The isothermal gas density is in hydrostatic equilibrium with the dark matter potential, 
\begin{equation}
    \frac{c_\mathrm{x}^2}{\gamma} \frac{d \ln{\rho_g}}{dr} = - \frac{G M(r)}{r^2}
    \label{eqn:hydrostatic-equilibrium-ch3}
\end{equation}
where $c_\mathrm{x}$ is the sound speed, $\gamma = 5/3$ is the adiabatic index, and $M(r)$ is the mass distribution of the system.
Neglecting the effect of self-gravity, and assuming that dark matter haloes follow the universal profile derived by \citet{Navarro1997}, the gas density profile can be shown to follow \citep{Makino1998}
\begin{equation}
    \label{eqn:gas-density-profile}
    \rho_g(r) = \rho_{g0} e^{-27 b / 2} \left(1 + \frac{r}{r_s} \right)^{27 b / (2 r / r_s)}
\end{equation}
where $r_s$ is the scale radius, and $b$ is a scaling parameter given by

\begin{equation}
    \label{eqn:makinobparameter1998}
    b(M) \equiv \frac{8 \pi G \gamma \delta_c(M) \rho_{c0} r_s^2}{27 c_\mathrm{x}^2}
\end{equation}
Here $G$ is the gravitational constant, $\delta_c$ is the characteristic density \citep{Navarro1997}, and $\rho_c$ is the critical density.

\begin{figure*}
    \includegraphics{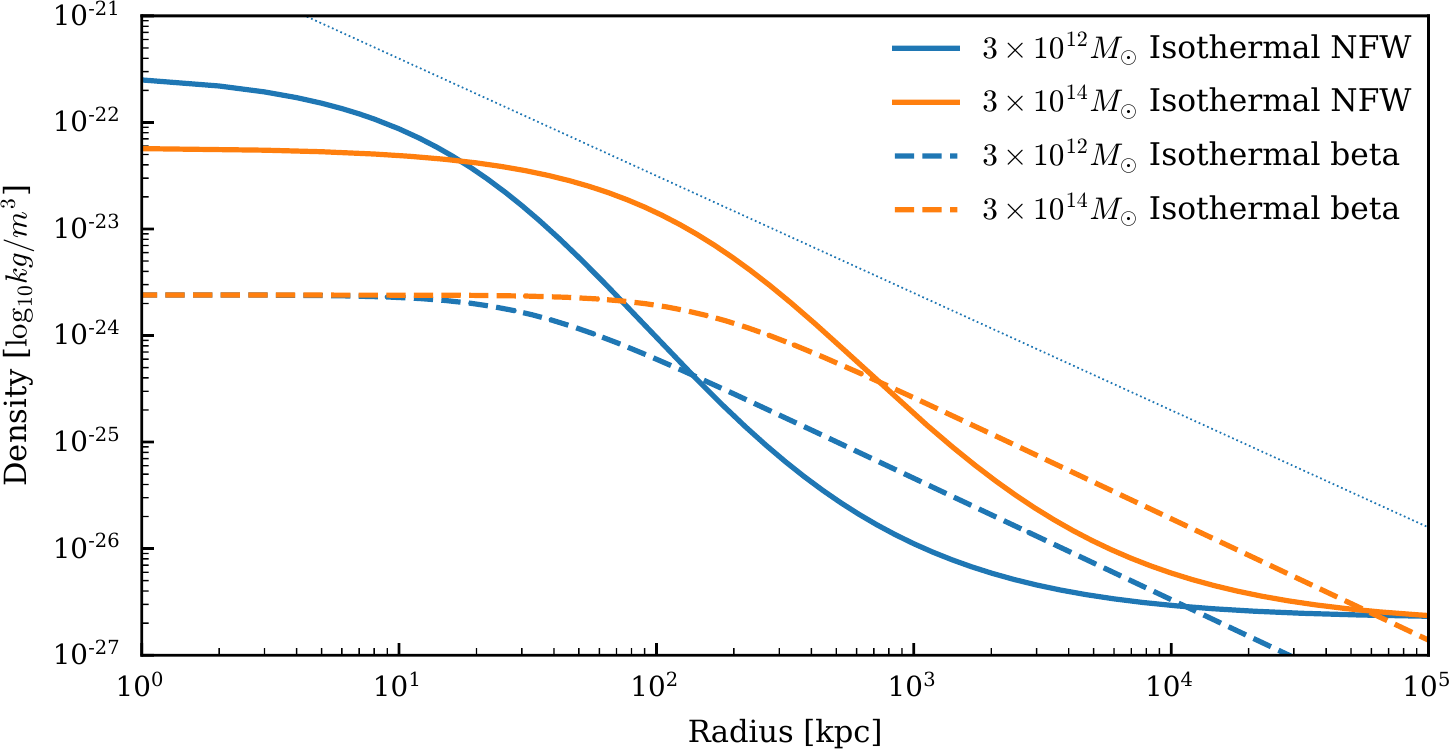}
    \caption{Isothermal NFW and beta density profiles for galaxy cluster- and group-like environments. The thin dotted line is a power-law with slope $\beta = -1.1$.}
    \label{fig:density-profiles}
\end{figure*}

The characteristic density parameter, $\delta_c$, requires calculating the concentration parameter for the dark matter halo mass, as
\begin{equation}
    \delta_c=\frac{\Delta}{3} \frac{c^3}{\ln(1+c)-c/(1+c)}
    \label{eqn:delta-c}
\end{equation}
We use the \citet{Klypin2016} concentration model for a relaxed dark matter halo,
\begin{equation}
    c(M) = c_0 \left( \frac{M}{10^{12} h^{-1} M_\odot} \right)^{- \kappa} \left[ 1 + \left( \frac{M}{M_0} \right)^{0.4} \right]
    \label{eqn:concentration-klypin-relaxed}
\end{equation}
where $c_0 = 7.75$, $\kappa = 0.100$, and $M_0 / 10^{12} h^{-1} M_\odot = 4.5\times10^5$.
Two different dark matter halo masses are simulated: $M_\mathrm{halo} = 3 \times 10^{14}{M_\odot}$, representative of a cluster environment; and $M_\mathrm{halo} = 3 \times 10^{12}{M_\odot}$, representative of a poor group.
\autoref{fig:density-profiles} shows the resulting isothermal NFW and beta profiles for these two dark matter halo masses.

The simulations do not include cooling, and the gas density profile used assumes that the environment is isothermal.
On the one hand, isothermality may be a reasonable assumption for clusters with efficient thermal conduction \citep{Narayan2001}.
On the other hand, observations \citep{Vikhlinin2006} show that ICM temperature may change by a factor of 2-3 over a virial radius.
Compared with non-isothermal clusters, our assumed density profiles exhibit higher core densities.

We also perform comparison runs in an isothermal beta profile \citep{King1962} of the same mass,

\begin{equation}
    n = n_0 \left[ 1 + \left( \frac{r}{r_c} \right)^2 \right]^{-3 \beta / 2}
    \label{eqn:king-profile}
\end{equation}
which has a corresponding gravitational potential described by \citet{Krause2005}.

The central cooling time $t_\mathrm{cool}$ for the isothermal cluster (temperature of $3.4\times10^{7}\,\mathrm{K}$) and group (temperature of $1.6\times10^{6}\,\mathrm{K}$) environments are on the order of $1.5\,\mathrm{Gyr}$ and $50\,\mathrm{Myr}$ respectively, assuming a representative core density of $\rho_g = 10^{-23}\,\mathrm{kg\,m}^{-3}$ and a metallicity of $Z=-1.0$ \citep{Sutherland1993}.

\subsection{Simulation runs}

\begin{table*}
    \centering
    \begin{tabular}{lrrrrrrrrrl}
        \hline
        Type & $M_\mathrm{halo}$ & $r_\mathrm{s,c}$ ($\mathrm{kpc}$) & $\rho_0$ ($\mathrm{kg\,m}^{-3}$) & $c_\mathrm{x}$ ($\mathrm{km\,s}^{-1}$) & $\rho_\mathrm{j}$ ($\mathrm{kg\,m}^{-3}$) & $L_1$ (kpc) & $\tau$ (Myr) & $\mathcal{M}_\mathrm{x}$ & $n$ & Run code \\
        \hline
        Isothermal NFW  & $3 \times 10^{12}$ & $42.8$ & $2.9\times 10^{-22}$ & $191$ & $1.3\times10^{-24}$ & $1.64\times10^{0}$  & $8.35\times10^{0}$  & 25  & 1 & {m12.5-M25-n1}          \\
                        &                    &        &                      &       & &                     &                     &     & 2 & {m12.5-M25-n2}          \\
                        &                    &        &                      &       & &                    &                     &     & 3 & {m12.5-M25-n3}          \\
                        &                    &        &                      &       & &                    &                     &     & 4 & {m12.5-M25-n4}          \\ [0.25em]

                        &                    &        &                      &       & &                    &                     & 25  & 1 & {m12.5-M25-n1-high-res} \\
                        &                    &        &                      &       & $4.1\times10^{-24}$ & $3.15\times10^{-1}$ & $1.61\times10^{0}$  & 75  & 1 & {m12.5-M75-n1}          \\
                        &                    &        &                      &       & $1.1\times10^{-23}$ & $7.23\times10^{-2}$ & $3.70\times10^{-1}$ & 200 & 1 & {m12.5-M200-n1}         \\ [0.75em]
                                                                                      
                        & $3 \times 10^{14}$ & $303$  & $5.8\times 10^{-23}$ & $888$ & $6.7\times10^{-24}$ & $3.65\times10^{-1}$ & $4.01\times10^{-1}$ & 25  & 1 & {m14.5-M25-n1}          \\
                        &                    &        &                      &       & &                    &                     &     & 2 & {m14.5-M25-n2}          \\
                        &                    &        &                      &       & &                    &                     &     & 3 & {m14.5-M25-n3}          \\
                        &                    &        &                      &       & &                    &                     &     & 4 & {m14.5-M25-n4}          \\ [0.25em]

                        &                    &        &                      &       & &                    &                     & 25  & 1 & {m14.5-M25-n1-high-res} \\
                        &                    &        &                      &       & $1.7\times10^{-23}$ & $7.02\times10^{-2}$ & $7.72\times10^{-2}$ & 75  & 1 & {m14.5-M75-n1}          \\
                        &                    &        &                      &       & $4.4\times10^{-23}$ & $1.61\times10^{-2}$ & $1.77\times10^{-2}$ & 200 & 1 & {m14.5-M200-n1}         \\ [0.75em]

        Isothermal beta & $3 \times 10^{12}$ & $30.9$ & $2.4\times 10^{-24}$ & $191$ & $1.3\times10^{-25}$ & $1.78\times10^{1}$  & $9.12\times10^{1}$  & 25  & 1 & {m12.5-M25-n1-beta}     \\
                        &                    &        &                      &       & &                    &                     &     & 4 & {m12.5-M25-n4-beta}     \\
                        & $3 \times 10^{14}$ & $144$  & $2.4\times 10^{-24}$ & $888$ & $1.2\times10^{-26}$ & $1.78\times10^{0}$  & $1.96\times10^{0}$  & 25  & 1 & {m14.5-M25-n1-beta}     \\
                        &                    &        &                      &       & &                    &                     &     & 4 & {m14.5-M25-n4-beta}     \\
        \hline
    \end{tabular}
    \captionsetup{justification=centering}
    \caption{Parameters for the simulation runs. $r_\mathrm{s,c}$ is the scale (for NFW) or core (for beta profile) radius, $M_\mathrm{halo}$ is the dark matter halo mass, $\rho_0$ is the central density, $c_\mathrm{x}$ is the sound speed, $\rho_\mathrm{j}$ is the collimated jet density, $L_1$ is the simulation length scale, $\tau$ is the simulation time scale, $\mathcal{M}_\mathrm{x}$ is the external Mach number, and $n$ is the number of outbursts. For all runs $Q=10^{37}\,\mathrm{W}$, $t_\mathrm{on} = 40\,\mathrm{Myr}$, $t_\mathrm{off} = 160\,\mathrm{Myr}$.}
    \label{tbl:simulation-runs}
\end{table*}

The aim of this work is to quantify the effects of environment and details of energy injection on the jet-environment interaction.
We propagate jets of the same kinetic power in different environments, as well as injecting the same total energy in either single or multiple bursts.
Our simulation runs are shown in \autoref{tbl:simulation-runs}.
For all simulations, the jet power is $Q_\mathrm{jet} = 10^{37}\,\mathrm{W}$, the total jet active time is $t_\mathrm{on,total} = 40\,\mathrm{Myr}$, the total jet quiescent time is $t_\mathrm{off,total} = 160\,\mathrm{Myr}$, the half-opening angle is $\theta_\mathrm{jet} = 15^\circ$, and the redshift is $z=0$.

The standard simulations are carried out in an environment given by the isothermal NFW gas density profile, with two different dark matter halo masses: $M_\mathrm{halo} = 3 \times 10^{14}{M_\odot}$ and $M_\mathrm{halo} = 3 \times 10^{12}{M_\odot}$.
One, two, three and four outburst jets are simulated, with the total jet active time of $40\,\mathrm{Myr}$, divided equally between the number of outbursts.
Each simulation is given a short run code which describes the key parameters of that simulation; for example `m14.5-M25-n4' corresponds to a simulation with $M_\mathrm{halo}=3 \times 10^{14}\,\mathrm{M}_\odot$, $M_\mathrm{x}=25$, and an outburst count of $n=4$.
The following additional simulations are carried out: four simulations in an environment given by the isothermal beta gas density profile \citep{King1962}; two high resolution Mach 25 simulations in the isothermal NFW environment; and four high Mach number (75, 200) simulations also in the isothermal NFW environment.
Due to the extra computational time required for the high Mach number and high resolution simulations, these were only run out to between $20$ and $40\,\mathrm{Myr}$.

The low resolution, Mach 25 simulations were run on the Tasmanian Partnership for Advanced Computing (TPAC) \textit{vortex} cluster at the University of Tasmania, using between 16 and 64 Xeon CPU cores.
Each of the low resolution, Mach 25 runs took between 160 (poor group) and 1,920 (cluster) CPU hours.
The high resolution and high Mach number simulations were run on the newer TPAC \textit{kunanyi} cluster, using between 1000 and 4000 Xeon CPU cores.
The Mach 200 simulation in the cluster environment (our most computationally intensive simulation) took approximately 1 million CPU hours to reach $t=40\,\mathrm{Myr}$.

\section{Environmental Dependence}\label{sec:env-dependence}

The effect of different environments on jet evolution is studied in this section, using simulations where the same radio jet is injected into galaxy cluster- and poor group-like environments.

We begin by comparing a single outburst radio jet with an active time of $40\,\mathrm{Myr}$, corresponding to run codes {m14.5-M25-n1} and {m12.5-M25-n1} for the cluster and poor group environments respectively.
The injected jet in both simulations has the same kinetic power, external Mach number, and opening angle.

In \autoref{sec:env-jet-dynamics} the morphological changes between radio jets collimating in different environments are examined.
Next in \autoref{sec:env-pd} we estimate the radio luminosity of jet inflated lobes to produce P-D tracks, a standard tool for studying radio galaxies.
We explore how simulated radio sources would appear when observed with a radio interferometer in \autoref{sec:env-sb}, by creating synthetic radio distributions of simulated sources.

\subsection{Jet dynamics}\label{sec:env-jet-dynamics}

\begin{figure*}
    \includegraphics{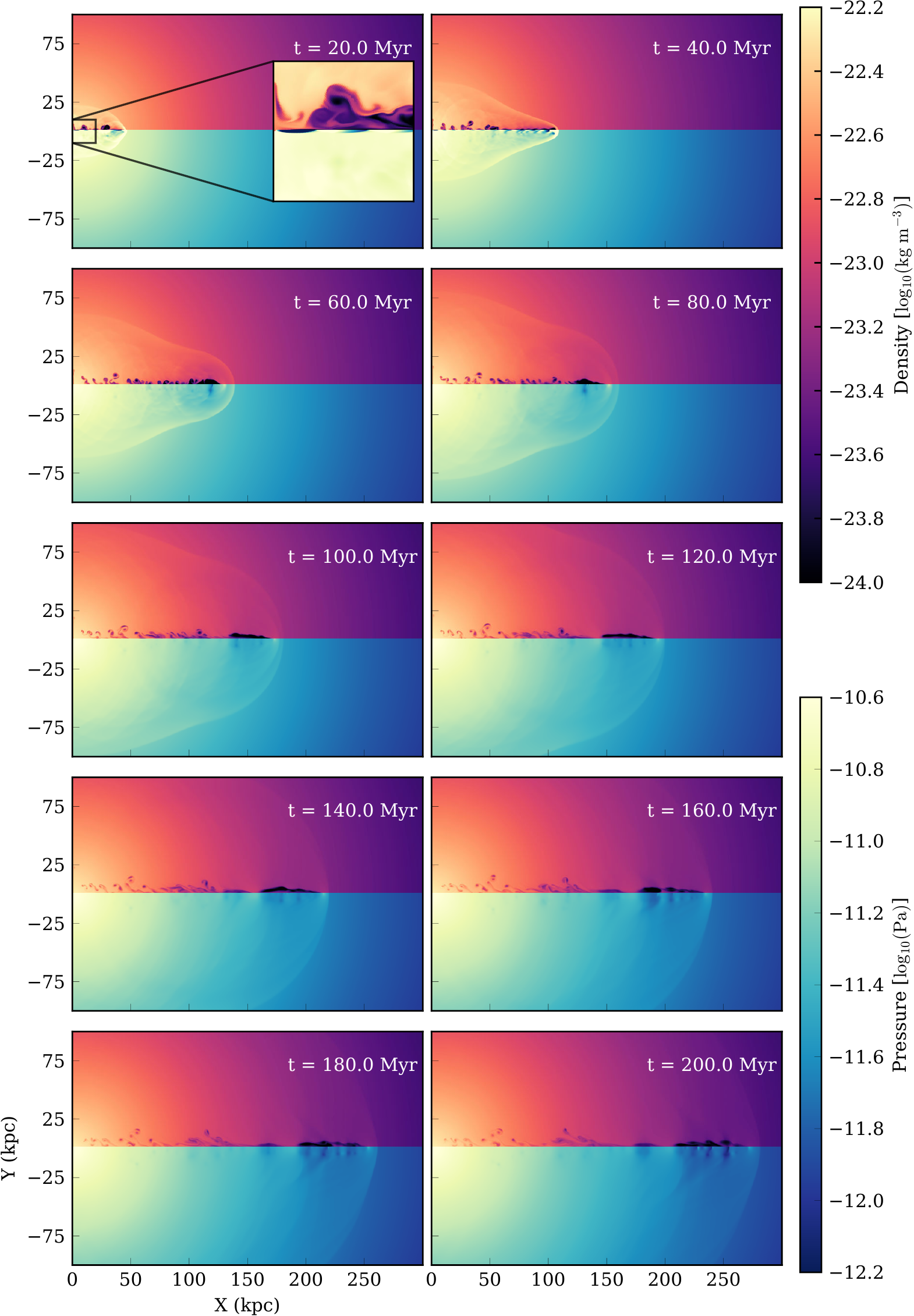}
    \caption{Density (upper) and pressure (lower) maps for a single outburst jet in the cluster environment, run m14.5-M25-n1, at ten different times. The inset in the upper-left panel shows the inner $20\,\mathrm{kpc}$ region.}
    \label{fig:dp-n1-cluster}
\end{figure*}

\begin{figure*}
    \includegraphics{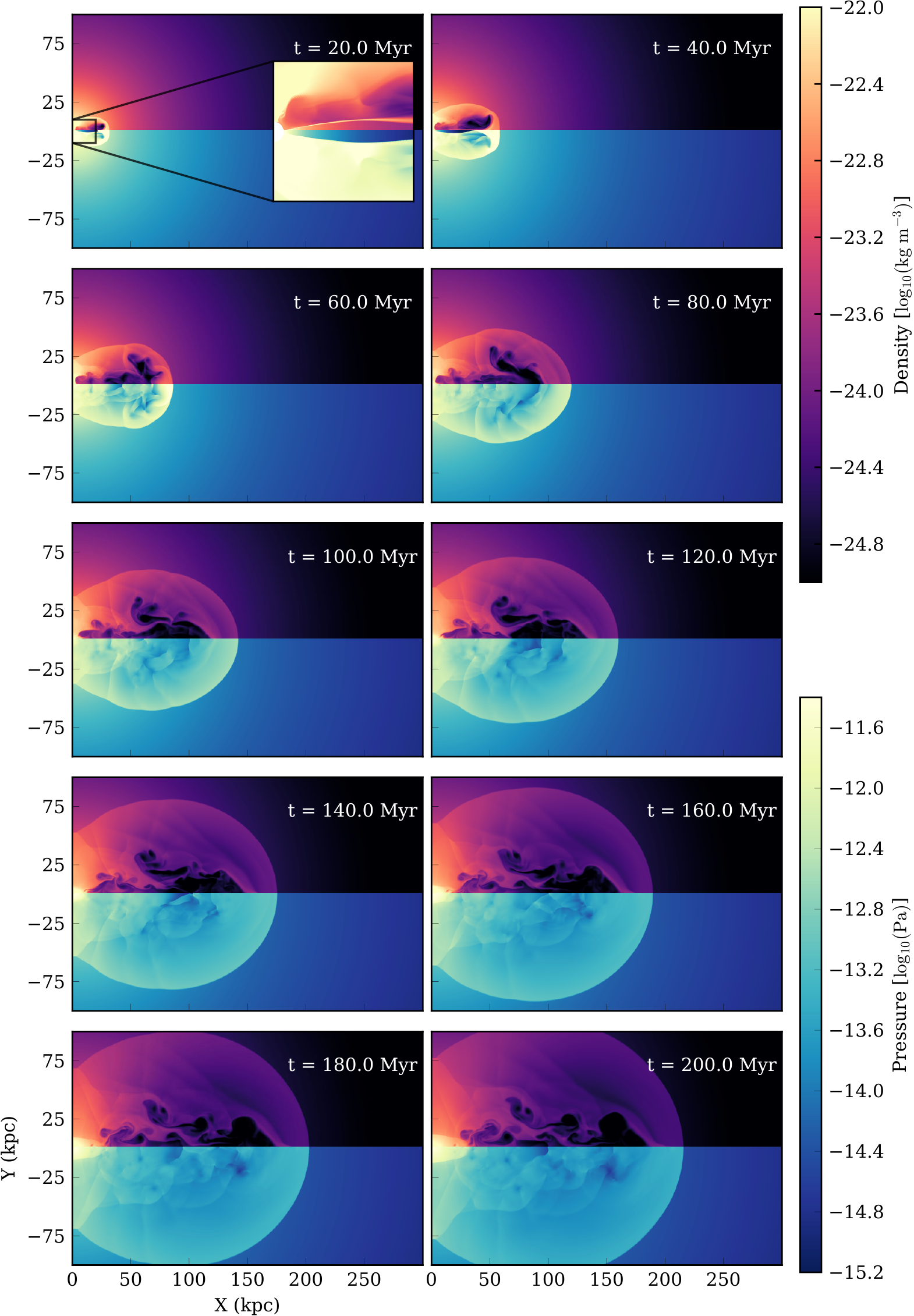}
    \caption{Density (upper) and pressure (lower) maps for a single outburst jet in the poor group environment, run m12.5-M25-n1, at ten different times. The inset in the upper-left panel shows the inner $20\,\mathrm{kpc}$ region.}
    \label{fig:dp-n1-poor-group}
\end{figure*}

\begin{figure*}
    \includegraphics{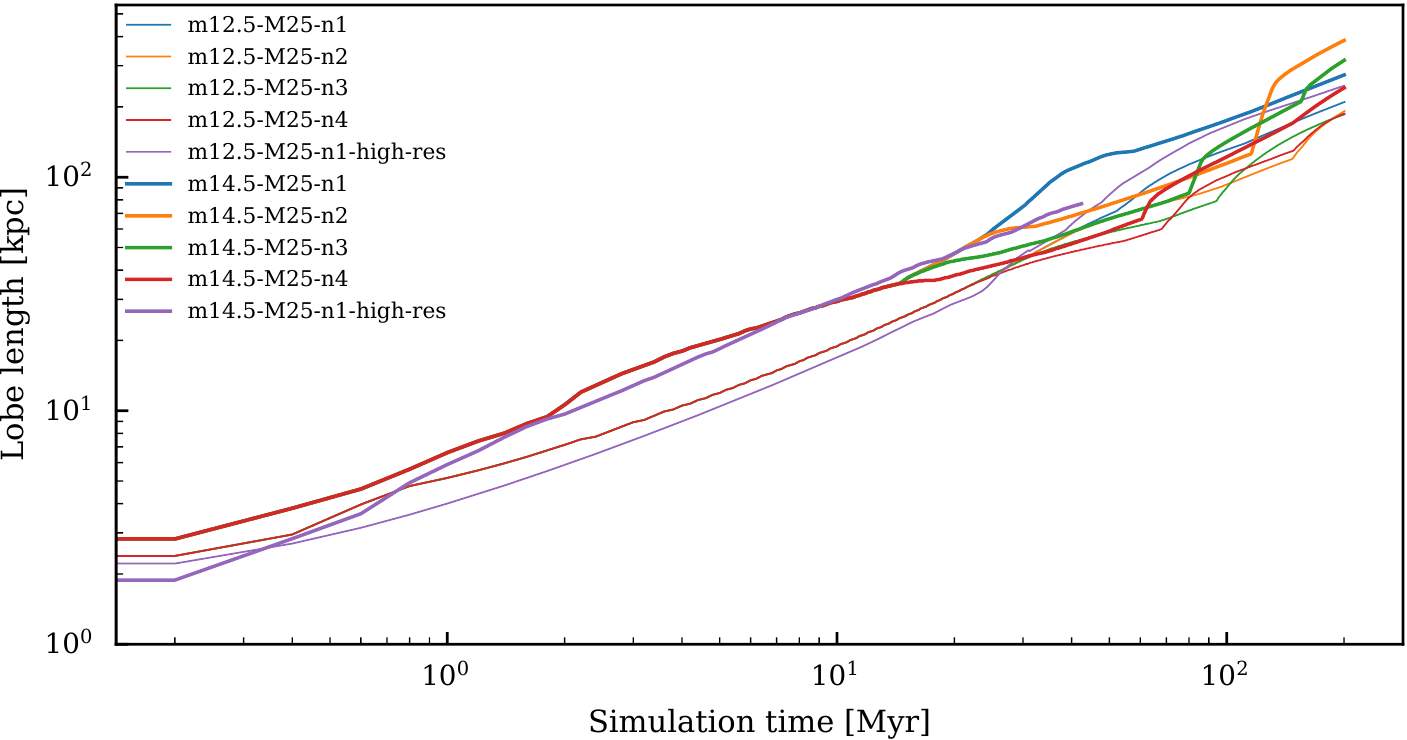}
    \caption{Evolution of lobe length for simulations in the cluster and poor group isothermal NFW environments.}
    \label{fig:ll-all}
\end{figure*}

\begin{figure*}
    \includegraphics{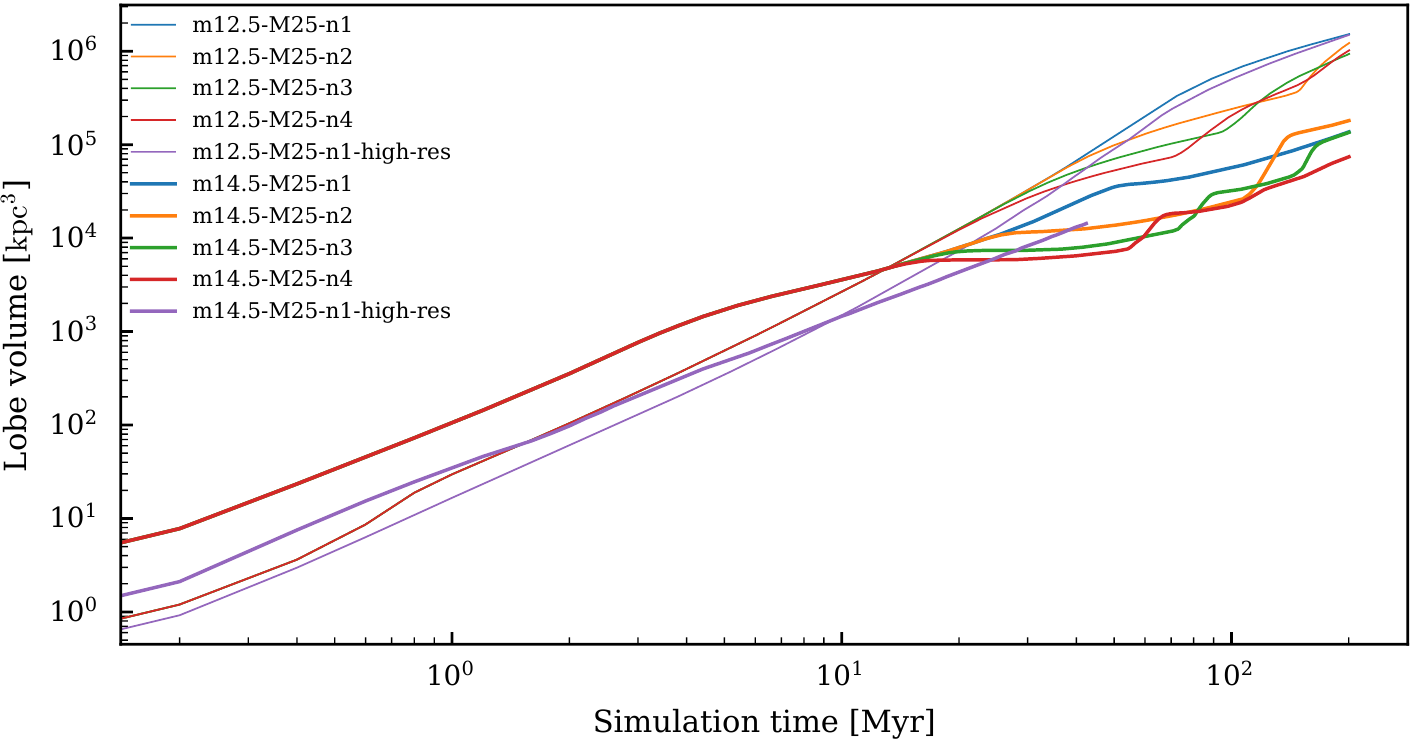}
    \caption{Evolution of lobe volume for simulations in the cluster and poor group isothermal NFW environments.}
    \label{fig:lv-all}
\end{figure*}

The evolution of the single outburst jet in the cluster and poor group environments is shown in \hyperref[fig:dp-n1-cluster]{Figures \ref*{fig:dp-n1-cluster}} and \ref{fig:dp-n1-poor-group}.
Both simulations reproduce the following key features of FR II morphology: a collimated jet; hotspot; cocoon inflated by backflow; and the bow shock.
This indicates that the simulations accurately capture large-scale jet dynamics.

The jet is injected as a conical mass inflow, and therefore a recollimation shock is expected to occur \citep[e.g.][hereafter KA97]{Kaiser1997}.
For a conically expanding jet, the jet density drops as $\rho_\mathrm{j} \propto 1/r^2$, and the jet collimates when the external pressure $p_\mathrm{x}$ is comparable to the sideways ram-pressure of the conical jet, $\rho_\mathrm{j} v_\mathrm{j}^2 \sin^2 \theta$, where $v_\textrm{j}$ is the jet velocity and $\theta$ is the half-opening angle.
This will occur on the length scale of $L_{1a}$ (see \citealt{Krause2012} for details), the analytically expected recollimation length scale.

The effect of recollimation is most visible in the poor group environment, where it occurs around $Y=14\,\mathrm{kpc}$ (see inset in the $t=20\,\mathrm{Myr}$ panel of \autoref{fig:dp-n1-poor-group}).
This large recollimation distance results in a low jet density and a wide jet, and is in agreement with the approximate expected (assuming a constant-density environment) length scale for the poor group environment of $L_{1a}=14.8\,\mathrm{kpc}$.
The jet in the cluster collimates on a smaller length scale of $L_{1a}=3.3\,\mathrm{kpc}$ as shown in \autoref{fig:dp-n1-cluster}.
The larger $L_{1a}$ for the poor group is due to the lower external pressure $p_\textrm{x}$ compared to the cluster, indicative of systematic correlation in simulations between the collimated width, jet density and environment.
The smaller recollimation length scale for the same half-opening angle produces a narrower collimated jet beam.

The jet switches off at $40\,\mathrm{Myr}$, after which the cocoon transitions to a rising bubble phase.
The buoyancy velocity is initially close to the sound speed in the ambient gas, and approaches half the sound speed in later phases, consistent with analytical expectations \citep{Churazov2001}.

The jet in the cluster environment is expected to propagate faster than the jet in the poor group environment.
This might seem counter-intuitive, as for a collimated jet, one would predict that, for the same jet power, the jet in the lower density environment is faster.
However the narrower collimated jet in the cluster has a smaller working surface at the jet head over which the forward ram pressure of the jet is distributed, making it easier for the jet to ``punch through'' the gas.
The difference in propagation velocity can be measured by quantifying the lobe (cocoon) length as a function of time for both environments, as shown in \autoref{fig:ll-all}.
The lobe length is calculated as the grid cell with the largest radial distance from the core, along $\theta = 0$, that contains a jet tracer value above the chosen threshold value.
The tracer cutoff is chosen to be $5\times10^{-3}$ (for comparison, \citet{Hardcastle2013} used a tracer cutoff of $10^{-3}$), however the measured lobe lengths are not sensitive to the exact value used, provided it is significantly smaller than unity.
The jet in the cluster propagates faster and inflates a longer lobe compared to the jet in the poor group.
The effect of the jet switching off (at $t=40\,\mathrm{Myr}$ for the single outburst simulations) on the lobe length is visible for the cluster environment as an inflection in the lobe length curve, while no such feature is visible for the poor group environment.

\subsection{Evolutionary tracks}\label{sec:env-pd}

\begin{figure}
    \includegraphics{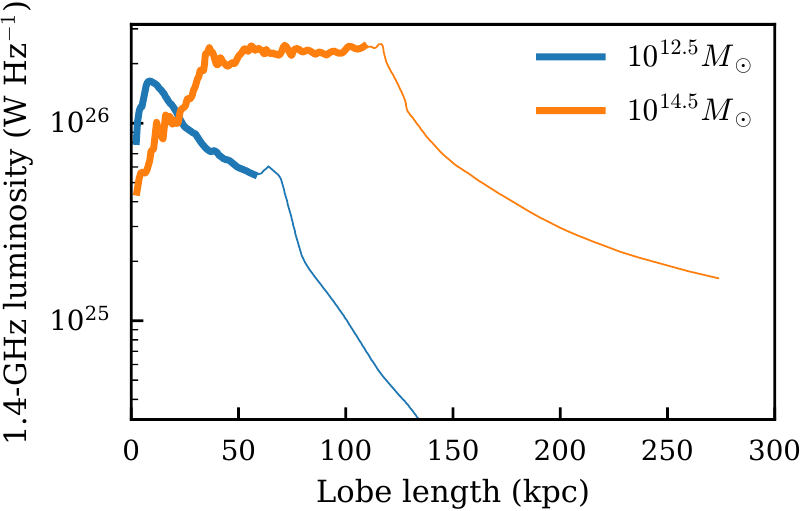}
    \caption{Size-luminosity diagram for the single outburst jet in the poor group and cluster environments. The thick lines are the active phase of the jet, while the thin lines are the passive phase. Note the linear scale for lobe length.}
    \label{fig:pd-n1}
\end{figure}

All the simulations carried out in this paper are purely hydrodynamical and do not contain the necessary physics (primarily magnetic fields) to completely calculate the synchrotron emissivity, as is done in, e.g., \citet{Jones1998}.
However, it is possible to calculate a lossless emissivity per unit volume by assuming that the pressure in the jet lobes is related to the electron energy density and magnetic field energy density.
This is the basis for radio source dynamical models (e.g.\@ those developed by \citet{Kaiser1997a,Kaiser2000,Willott1999,Kaiser2007,Shabala2008,Turner2015}), where the magnetic field energy density is typically assumed to be a constant fraction of the particle energy density.

The full derivation for the luminosity equation in terms of simulation quantities is given in \aref{app:synch-derivation}; the luminosity scaled to physical units ($\mathrm{W\ Hz}^{-1}$) is given by

\begin{equation}
    \label{eqn:luminosity-physical}
    L(\nu) = L_0 \left(\frac{\nu}{1\,\mathrm{GHz}} \right)^{ - \frac{q-1}{2}} \left( \frac{ p_0}{10^{-11}\,\mathrm{Pa}} \right)^{\frac{q+5}{4}} \left( \frac{L_1}{\mathrm{kpc}} \right)^3
\end{equation}

where $L_0=2.04\times 10^{23}\,\mathrm{W\,Hz}^{-1}$ is the coefficient for $L(\nu)$ scaled to $(L_1, p_0, \nu) = (1\,\mathrm{kpc}, 10^{-11}\,\mathrm{Pa}, 1\,\mathrm{GHz})$.

\citet{Kaiser1997a} developed a model for the evolution of a radio source through the P-D diagram, and showed that the shape of the track depends on the environment, with larger central densities corresponding to higher P-D tracks.
They modelled the gas density as a power law of the form $\rho_\mathrm{x} \propto r^{-\beta}$, where $r$ is the radius from the core.
An environment with $\beta \gtrsim 1.1$ will have a falling P-D track, while an environment with $\beta \lesssim 1.1$ will have a rising P-D track.

The total luminosity of the simulation is calculated by adding up the luminosity in each grid cell as given by eq.~(\ref{eqn:luminosity-physical}).
The ratio of magnetic to particle energy densities is taken to be $\eta = 0.1$, which is consistent with magnetic field strengths derived from observations \citep{Turner2018}.
The maximum and minimum energies are taken to correspond to Lorentz factors $\gamma_\textrm{min} = 10$ and $\gamma_\textrm{max}=10^5$ as in \citet{Hardcastle2013}.

The total luminosity is plotted against the lobe size in \autoref{fig:pd-n1} for the single outburst jet simulation in the cluster and poor group environments.
Here lobe length acts as a proxy for time (see \autoref{fig:ll-all}).

The environment greatly affects the total luminosity of jet-inflated structures.
The track of the jet in the poor group environment reproduces the peak of the evolutionary track and subsequent decline in overall luminosity at large jet sizes of the model developed by \citet{Kaiser1997a}.
The track of the jet in the cluster environment, on the other hand, continues to increase with increasing jet size as is expected due to the larger central region of approximately constant density.
The jet is still expanding into a significantly denser environment at large radii compared to the poor group, which increases the pressure and in turn increases the overall luminosity.
It's important to note that the luminosities calculated here neglect synchrotron and Inverse Compton losses (which are expected to not be significant for ages $\leq 40\,\mathrm{Myr}$), as well as re-acceleration of electrons at shocks.

\subsection{Surface brightness}\label{sec:env-sb}

\begin{figure*}
    \centering
    \begin{subfigure}[t]{0.5\textwidth}
        \centering
        \includegraphics{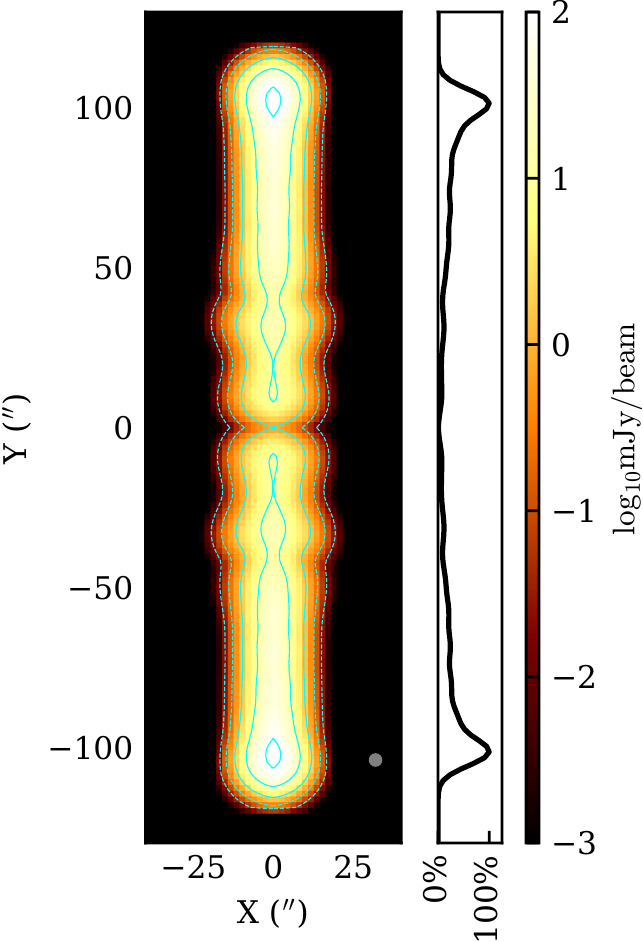}
        \caption{Cluster environment}
        \label{sfig:sb-n1-cluster}
    \end{subfigure}%
    ~
    \begin{subfigure}[t]{0.5\textwidth}
        \centering
        \includegraphics{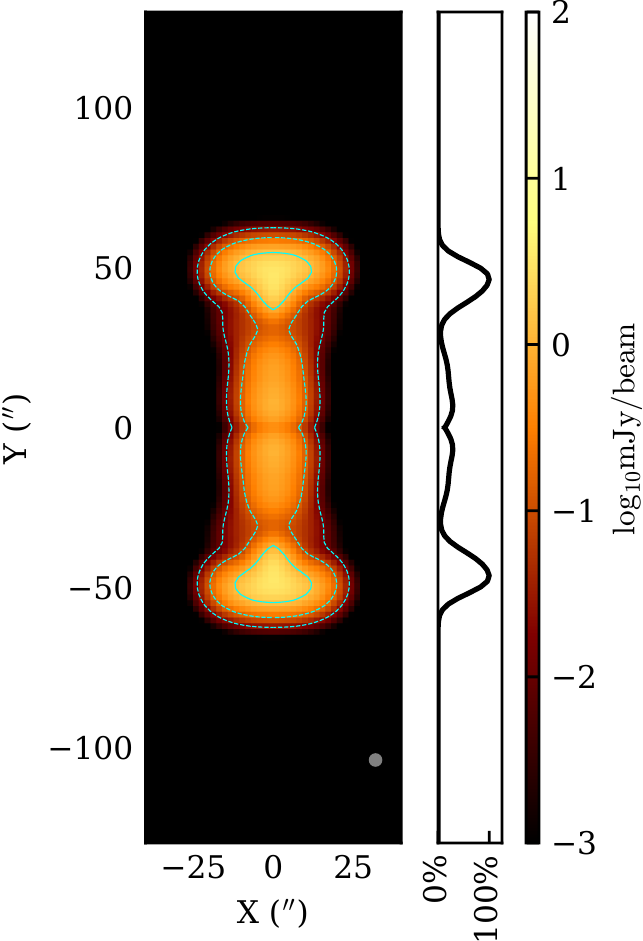}
        \caption{Poor group environment}
        \label{sfig:sb-n1-poor-group}
    \end{subfigure}%
    \caption{Surface brightness plots for single outburst jets at $t=40\,\mathrm{Myr}$. Contours are at $0.01$, $0.1$, $1$, $10$, $100$ $\mathrm{mJy}/\mathrm{beam}$, with dotted contours representing a surface brightness below the FIRST detection threshold. The surface brightness along the jet axis (scaled to the maximum surface brightness) is shown to the right of each plot. The ratio between kpc and arcsec at the chosen redshift $z=0.05$ is 1:1.}
\end{figure*}

\begin{figure*}
    \centering
    \begin{subfigure}[t]{0.5\textwidth}
        \centering
        \includegraphics{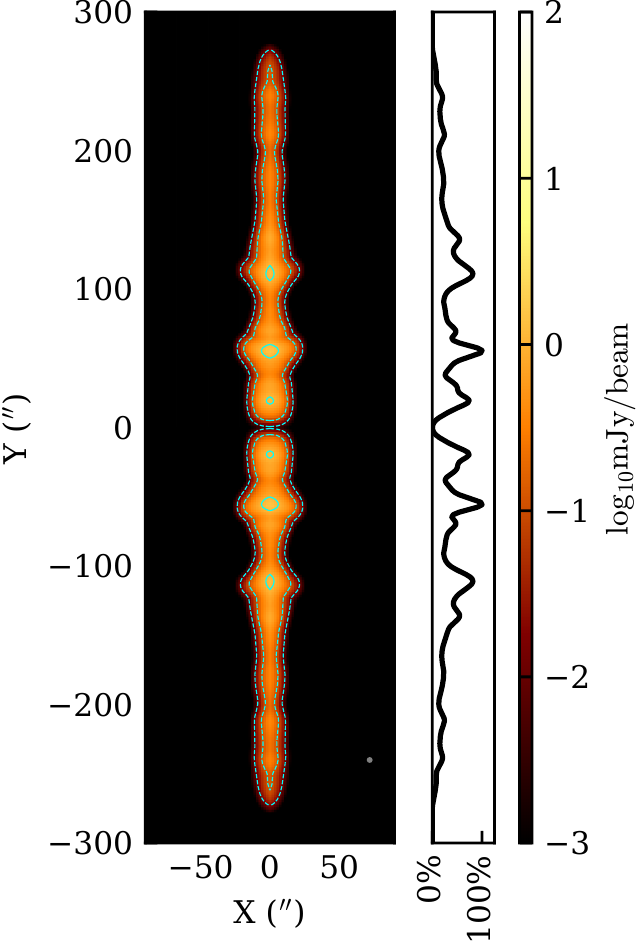}
        \caption{Cluster environment}
    \end{subfigure}%
    ~
    \begin{subfigure}[t]{0.5\textwidth}
        \centering
        \includegraphics{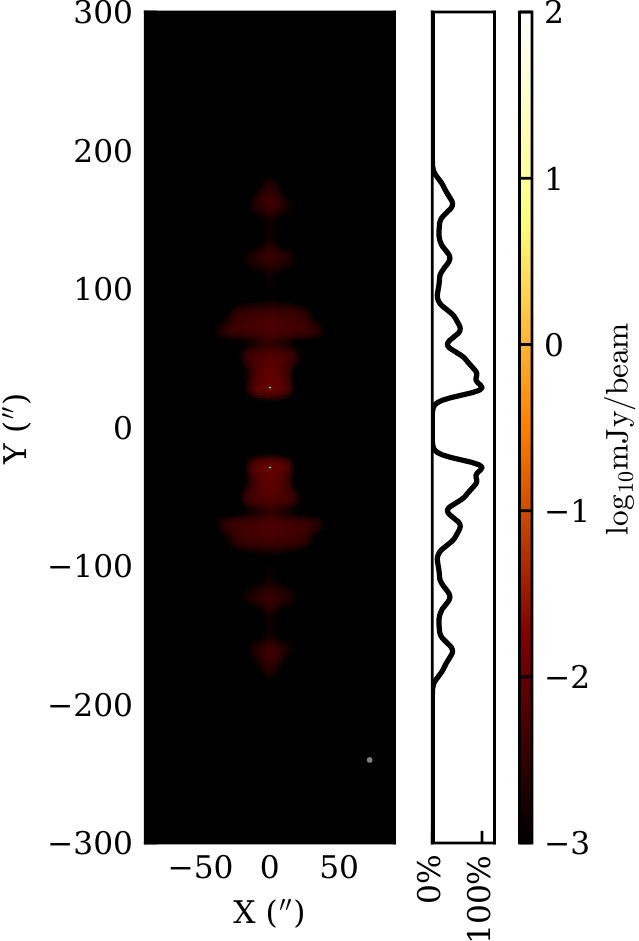}
        \caption{Poor group environment}
    \end{subfigure}%
    \caption{Surface brightness plots for single outburst jets at $t=200\,\mathrm{Myr}$. This is $160\,\mathrm{Myr}$ after the jet has switched off, and so radiative losses are no longer negligible. As these losses are not included in the radio model, the surface brightness maps are upper limits. Contours are at $0.01$, $0.1$, $1$, $10$, $100$ $\mathrm{mJy}/\mathrm{beam}$.}
    \label{fig:sb-n1-cluster-group}
\end{figure*}

The surface brightness of each grid cell is calculated from the luminosity equation given in Eq.~(\ref{eqn:luminosity-physical}), and is then weighted by the average value of the jet tracer.
This weighting process is necessary because each simulation cell may in principle contain both jet and non-jet material.
Weighting by the jet tracer value ensures that only the jet plasma is contributing to synchrotron emission.
We place our simulated radio galaxies at $z=0.05$ (corresponding to $1.0\,\mathrm{kpc}$ per arcsec) oriented in the plane of the sky, and ray-trace along sightlines through the projected three dimensional volume to obtain the total 2D surface brightness map.
Finally the simulated surface brightness is convolved with a 5'' (FWHM) beam, chosen to approximately match the FIRST survey \citep{Becker1995}.
\autoref{sfig:sb-n1-cluster} shows the surface brightness plot of the radio jet for $n=1$ in the cluster environment when the jet switches off at $t=40\,\mathrm{Myr}$, while \autoref{sfig:sb-n1-poor-group} shows the corresponding simulation for the poor group environment.
The surface brightness profile along the jet axis as a percentage of maximum surface brightness is shown to the right of the surface brightness map.

The general jet morphology is visible in the surface brightness distributions for both environments.
The recollimation of the jet is visible for both environments, and is more prominent with the jet in the poor group environment because of the larger recollimation length scale $L_{1a}$.
The hotspot shows as a region of high surface brightness at the head of the jet for both environments.
This confirms the FR II nature of the simulated radio sources.

However there is a large difference in the observed surface brightness distributions, due to the differing lobe morphology in the two environments.
The jet in the cluster has a higher overall surface brightness compared with the jet in the poor group, due to the higher pressure.
Taking $1\,\mathrm{mJy\,/\,beam}$ as the approximate $6-7 \sigma$ surface brightness detection threshold for the FIRST survey \citep{Becker1995}, a large part of the radio lobe for the poor group would not be detected.
Parts of the radio lobe inflated by the jet in the cluster environment would also fall below the detection limit if the source were moved out to higher redshift.
A large fraction of compact sources are observed in surveys (e.g. \citet{Sadler2014a}), and these are often found in poor environments \citep{Shabala2017,Shabala2018}, which is consistent with our simulations.
These simulations support the hypothesis of \citet{Shabala2017} that at least some of these sources may not be genuinely compact, but the lobes might fall (just) below the detection limit of current surveys, so that only the core (not simulated here) is detected.

\autoref{fig:sb-n1-cluster-group} shows how the surface brightness distributions for the single outburst jets in both environments evolve in the passive phase of the jet.
The $n=1$ simulation in the group environment produces a surface brightness distribution that contains extended lobe emission below the detection threshold of $\sim 1\,\mathrm{mJy\,/\,beam}$ for a FIRST-like survey.
Emission from this radio source would only be visible in observations sensitive to low surface brightness objects, such as the MWA GLEAM survey \citep{Hurley-Walker2015}.

\section{Intermittency}

\begin{figure*}
    \includegraphics{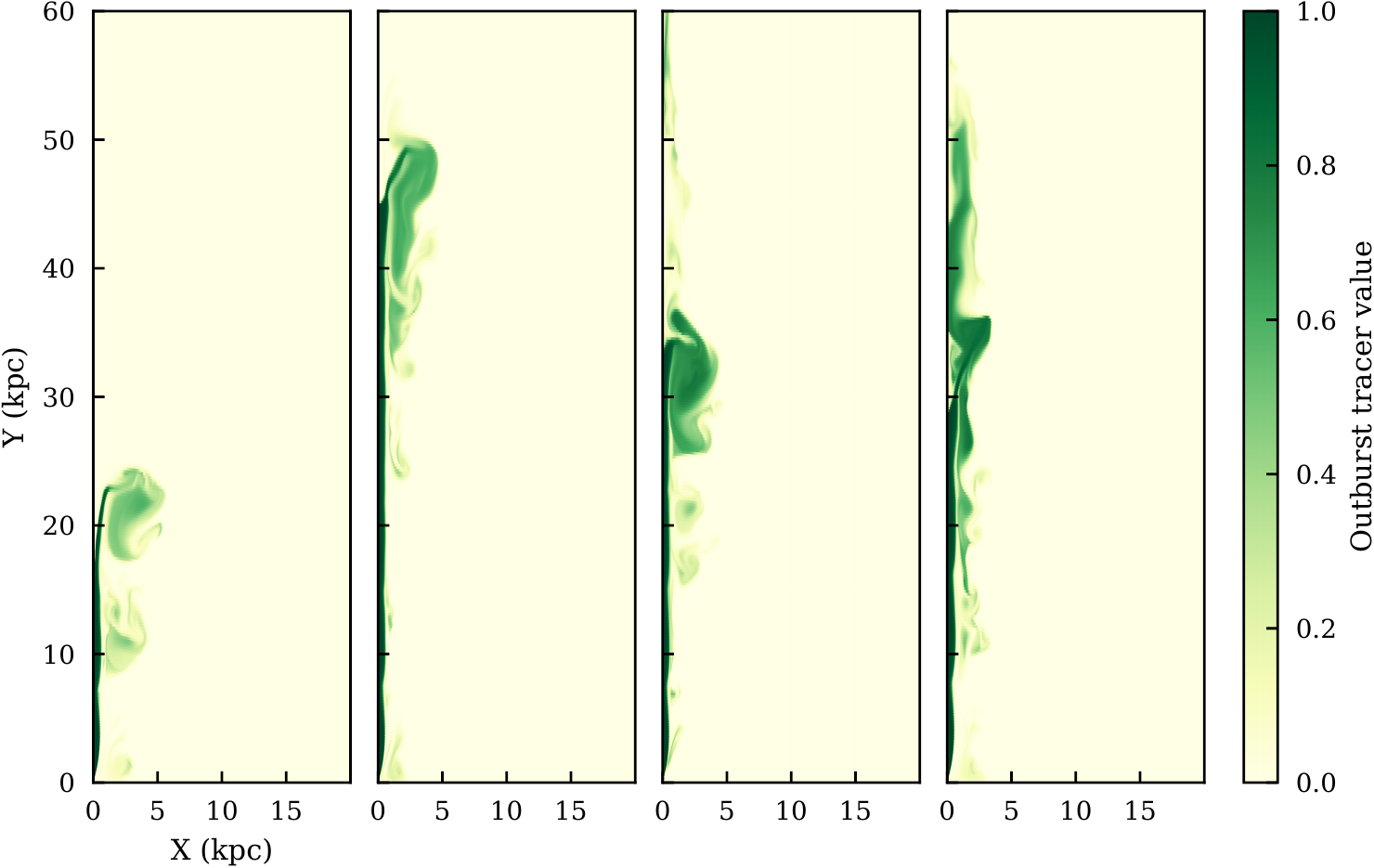}
    \caption{Outburst tracer map for a four outburst jet in the cluster environment, run m14.5-M25-n4. Each panel shows the simulation as an outburst is finishing, beginning with the first outburst in the leftmost panel. The times from left to right are $10\,\mathrm{Myr}$, $60\,\mathrm{Myr}$, $110\,\mathrm{Myr}$, $160\,\mathrm{Myr}$.}
    \label{fig:tm-cluster}
\end{figure*}

\begin{figure*}
    \includegraphics{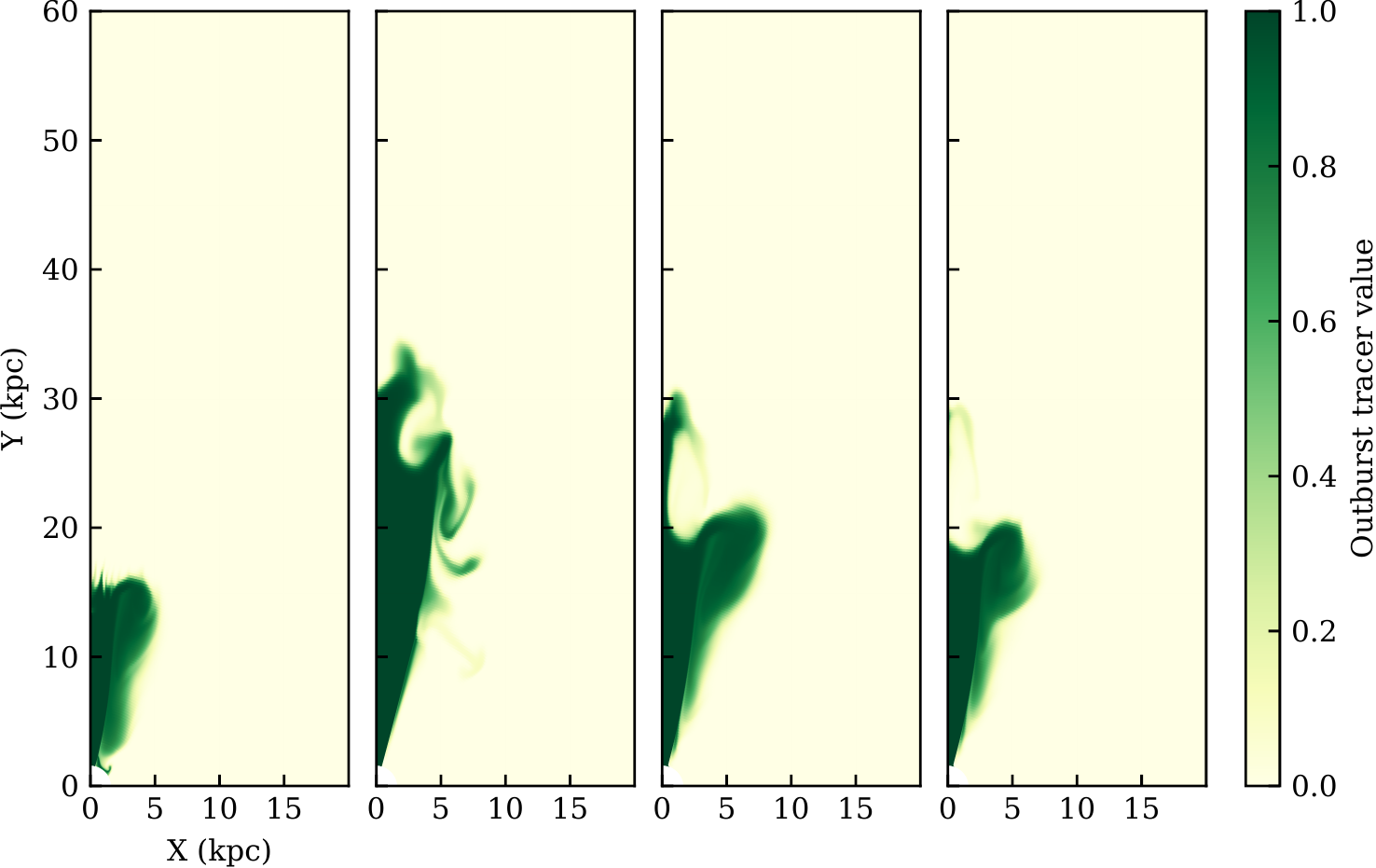}
    \caption{Outburst tracer map for a four outburst jet in the poor group environment, run m12.5-M25-n4. The panels and times are the same as \autoref{fig:tm-cluster}.}
    \label{fig:tm-poor-group}
\end{figure*}

\begin{figure*}
    \includegraphics{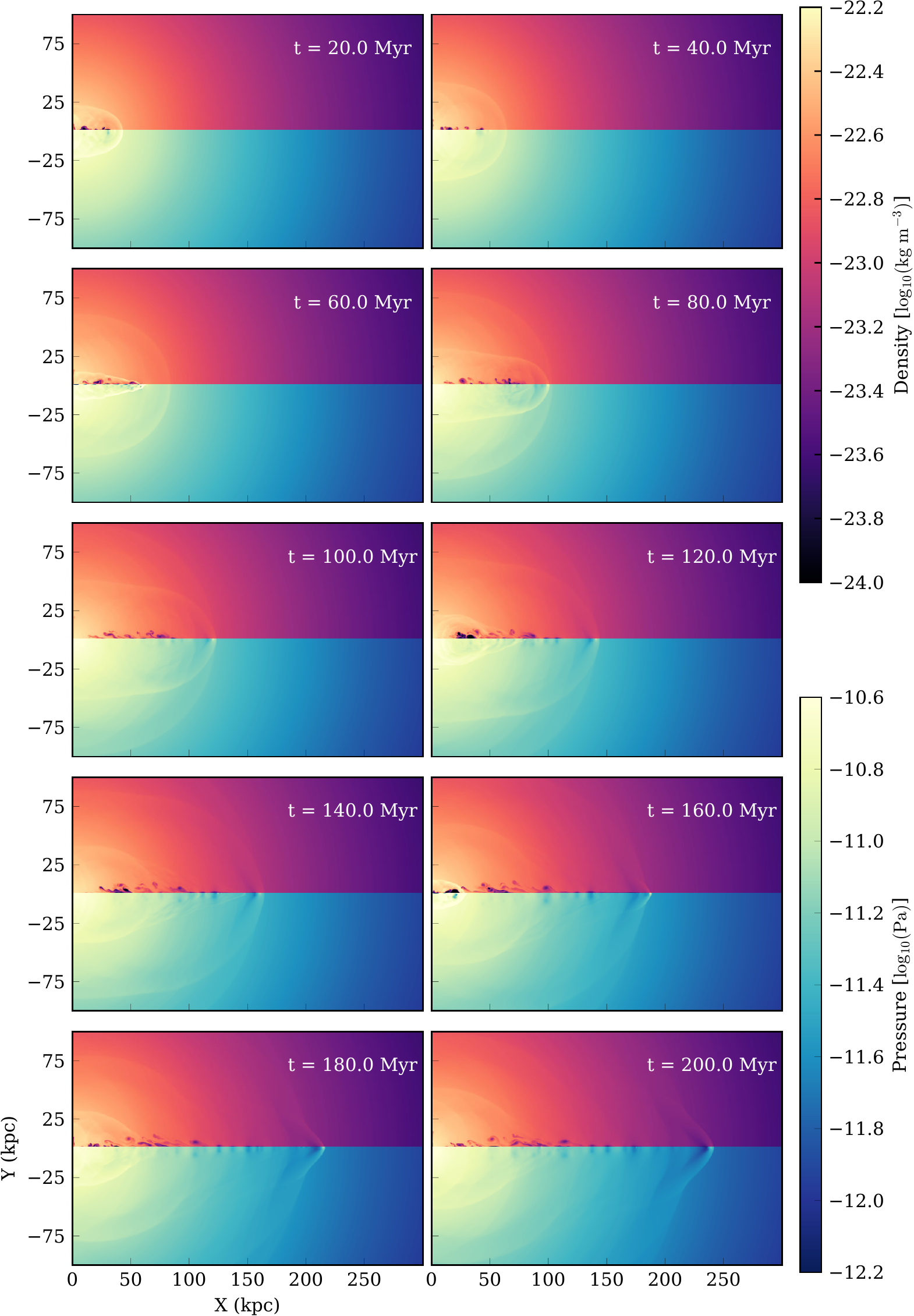}
    \caption{Density (upper) and pressure (lower) maps for a four outburst jet in the cluster environment, run m14.5-M25-n4, at ten different times.}
    \label{fig:dp-n4-cluster}
\end{figure*}

\begin{figure*}
    \includegraphics{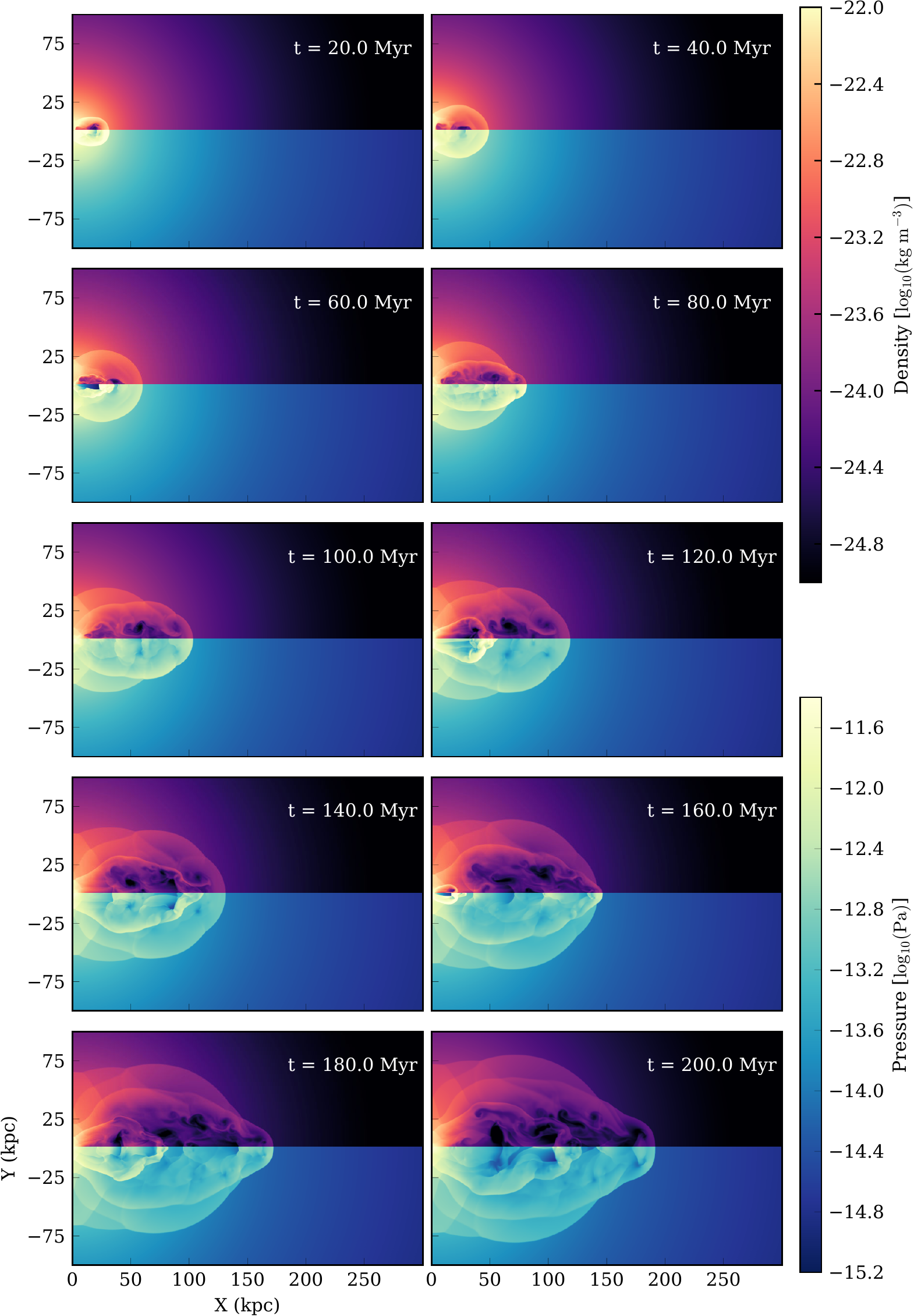}
    \caption{Density (upper) and pressure (lower) maps for a four outburst jet in the poor group environment, run m12.5-M25-n4, at ten different times.}
    \label{fig:dp-n4-poor-group}
\end{figure*}

We have shown how the environment affects jet dynamics and observable properties of radio sources in \autoref{sec:env-dependence}; now we present simulations of intermittent jets, and explore their effects on the same properties.
In this section we compare simulations of intermittent jets with the same duty cycle ($20$ per cent), but different number of outbursts.
The specific simulations compared are 1- and 4-outburst jets in both the cluster and poor group environments, simulations m14.5-M25-n1, m14.5-M25-n4, m12.5-M25-n1, and m12.5-M25-n4.
Simulations for n=2 and n=3 are also carried out, and their detectability is analysed.
We emphasize that in each simulation the jet delivers the same amount of energy over the same total amount of time ($10^{37}\,\mathrm{W}$ over $40\,\mathrm{Myr}$), and we only vary the intermittency of the jet injection.

We begin by examining how different numbers of outbursts affect the resulting jet dynamics in \autoref{sec:int-jet-dynamics}, and then show how this affects the observable properties of the source in \autoref{sec:int-radio}.

\subsection{Jet dynamics}\label{sec:int-jet-dynamics}

The basic Fanaroff-Riley type II morphology discussed in \autoref{sec:env-jet-dynamics} (such as the collimated jet, jet shock, hotspot, and bow shock) is reproduced for the $n=4$ outburst jet in both environments; however there are clear differences between the $n=1$ and $n=4$ jets.
We tag each outburst with a unique tracer particle, which is injected with a value of $1.0$.
The tracer value then decreases due to mixing with the environment, and can be used to track each outburst's jet material.
\autoref{fig:tm-cluster} and \autoref{fig:tm-poor-group} show tracer maps of the 4-outburst jet in the cluster and poor group environments respectively.
The jet outburst morphology is different at the end of each outburst, purely as a result of the preconditioned environment into which the jet is propagating.

The initial outburst evacuates a jet channel and material surrounding the core; this evacuation does not occur to the same degree for later outbursts.
Refilling of the jet channel and material surrounding the core occurs once the jet is switched off as inferred by \citet{Kaiser2000} from observations of double-double radio galaxies.
The jet of the second outburst is propagating into this partially refilled jet channel and so propagates faster.
By the time the third outburst begins the jet channel has almost completely refilled in the cluster environment (\autoref{fig:tm-cluster}) and so the jet propagates in a manner similar to the initial outburst.
In the poor group environment (\autoref{fig:tm-poor-group}) later outbursts collimate at larger radii than the initial outburst due to the preconditioned environment.

The time evolution for the 4-outburst jet in the cluster and poor group environments is shown as density and pressure maps in \autoref{fig:dp-n4-cluster} and \autoref{fig:dp-n4-poor-group} respectively.
Each outburst produces a corresponding bow shock, and these are visible in both environments.
The later bow shocks overtake the previous ones in the axial direction for both environments, as well as overtaking in the transverse direction for the poor group environment.
Subsequent jet inflated cocoons overtake previous ones due to the partially refilled old jet channel, through which the new plasma flows faster.

At a given age, the length of the radio lobe in the poor group environment is bounded by the results of the $n=1$ simulation.
This implies that multiple outbursts in the poor group environment are less effective at expanding the radio source than a single outburst of the same duration.
This is likely because of the larger collimated width for later outbursts shown in \autoref{fig:tm-poor-group} spreads jet momentum over a larger working surface.
A similar situation is seen when comparing the $n=1$ and $n=4$ simulations in the cluster environment; again multiple outbursts are less effective at expanding the radio source than a single outburst of the same duration.

We defer a detailed discussion of implications for AGN feedback in group and cluster environments to a future paper, and proceed with a discussion of the detectability of radio lobes inflated by multiple outbursts.

\subsection{Evolutionary tracks and radio dectability}\label{sec:int-radio}

\begin{figure}
    \includegraphics{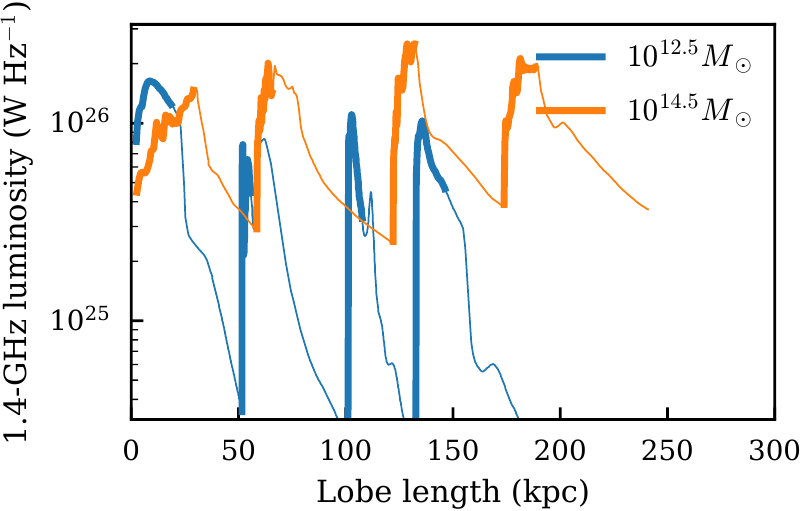}
    \caption{Size-luminosity diagram for the 4-outburst jet in the poor group and cluster environments. The thick lines are the active phases of the jet, while the thin lines are the passive phases.}
    \label{fig:pd-n4}
\end{figure}

\begin{figure*}
    \includegraphics{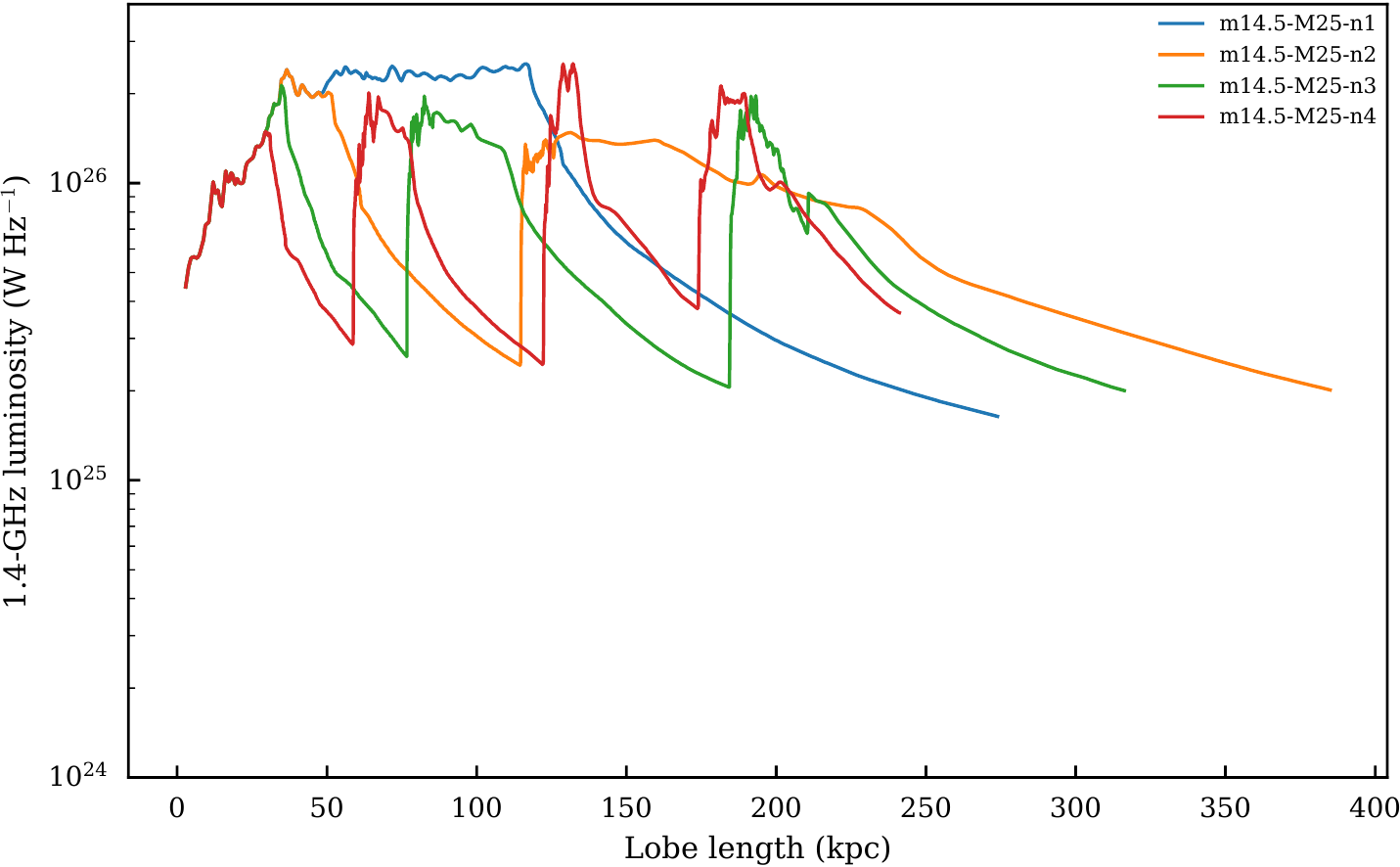}
    \caption{Size-luminosity diagram for 1, 2, 3, and 4 outbursts in the cluster isothermal NFW environment.}
    \label{fig:pd-c}
\end{figure*}

\begin{figure*}
    \includegraphics{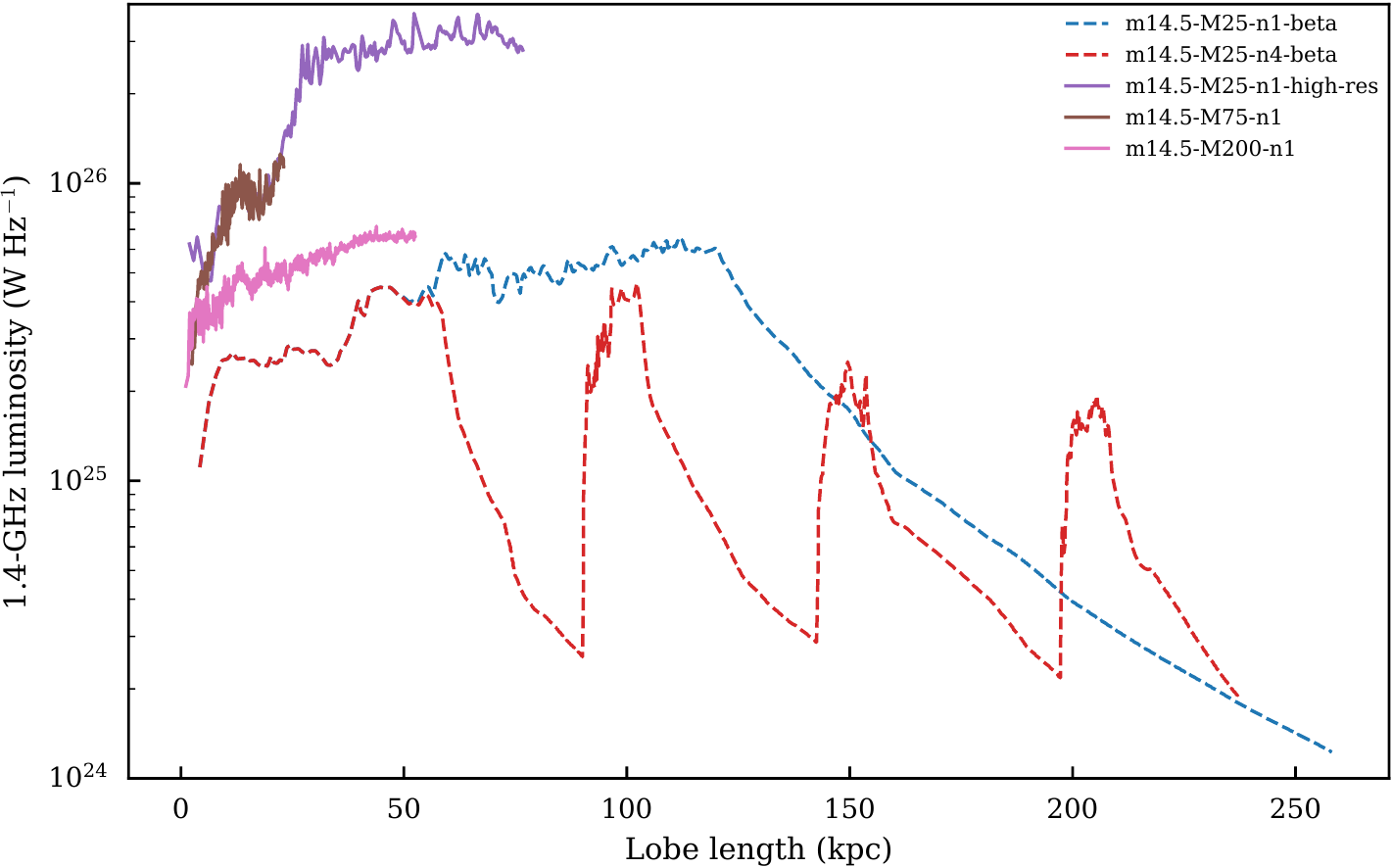}
    \caption{Size-luminosity diagram for supplementary (high resolution and high Mach number) simulations in the cluster environment.}
    \label{fig:pd-c-extra}
\end{figure*}

\begin{figure*}
    \includegraphics{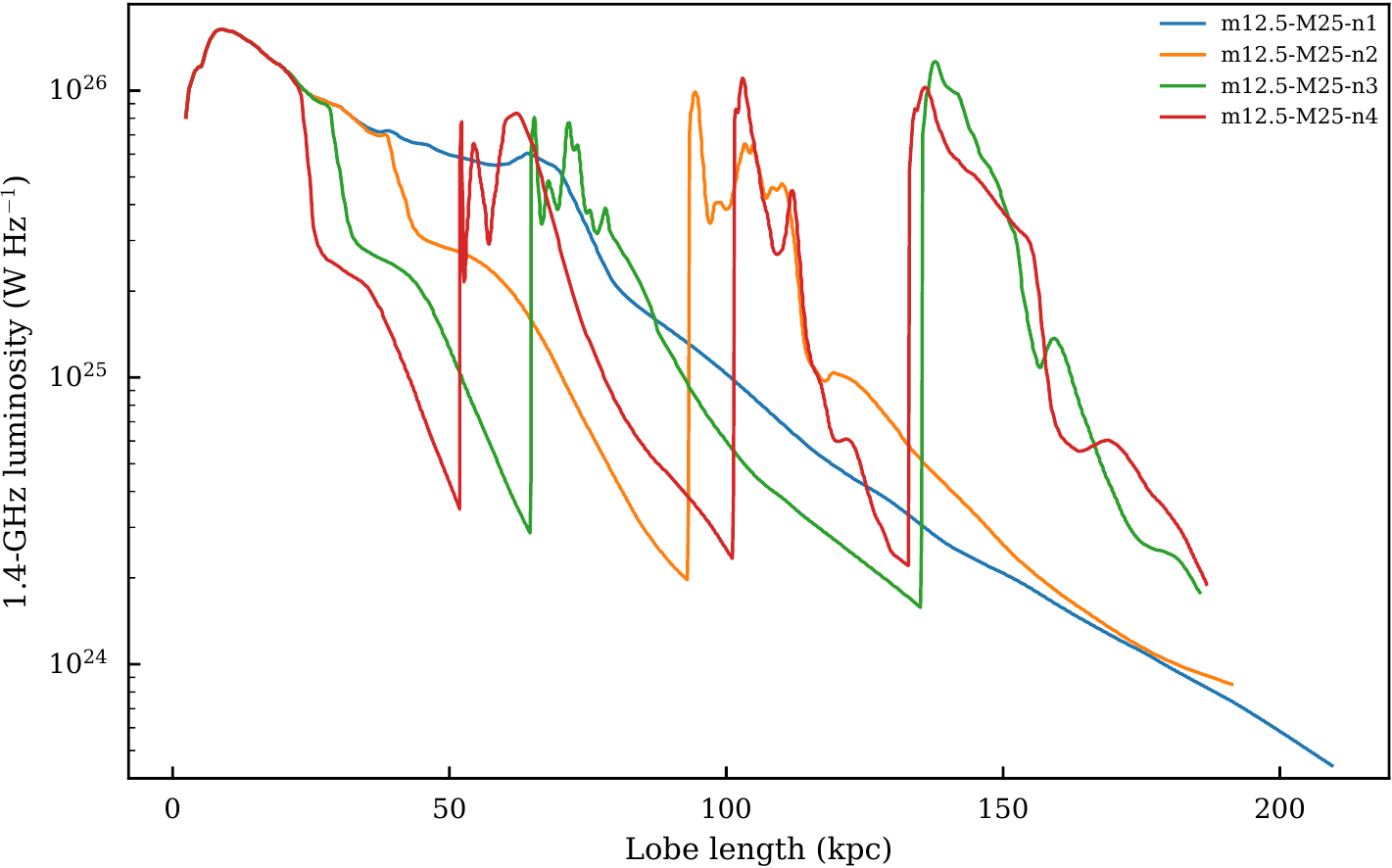}
    \caption{Size-luminosity diagram for 1, 2, 3, and 4 outbursts in the poor group isothermal NFW environment.}
    \label{fig:pd-pg}
\end{figure*}

\begin{figure*}
    \includegraphics{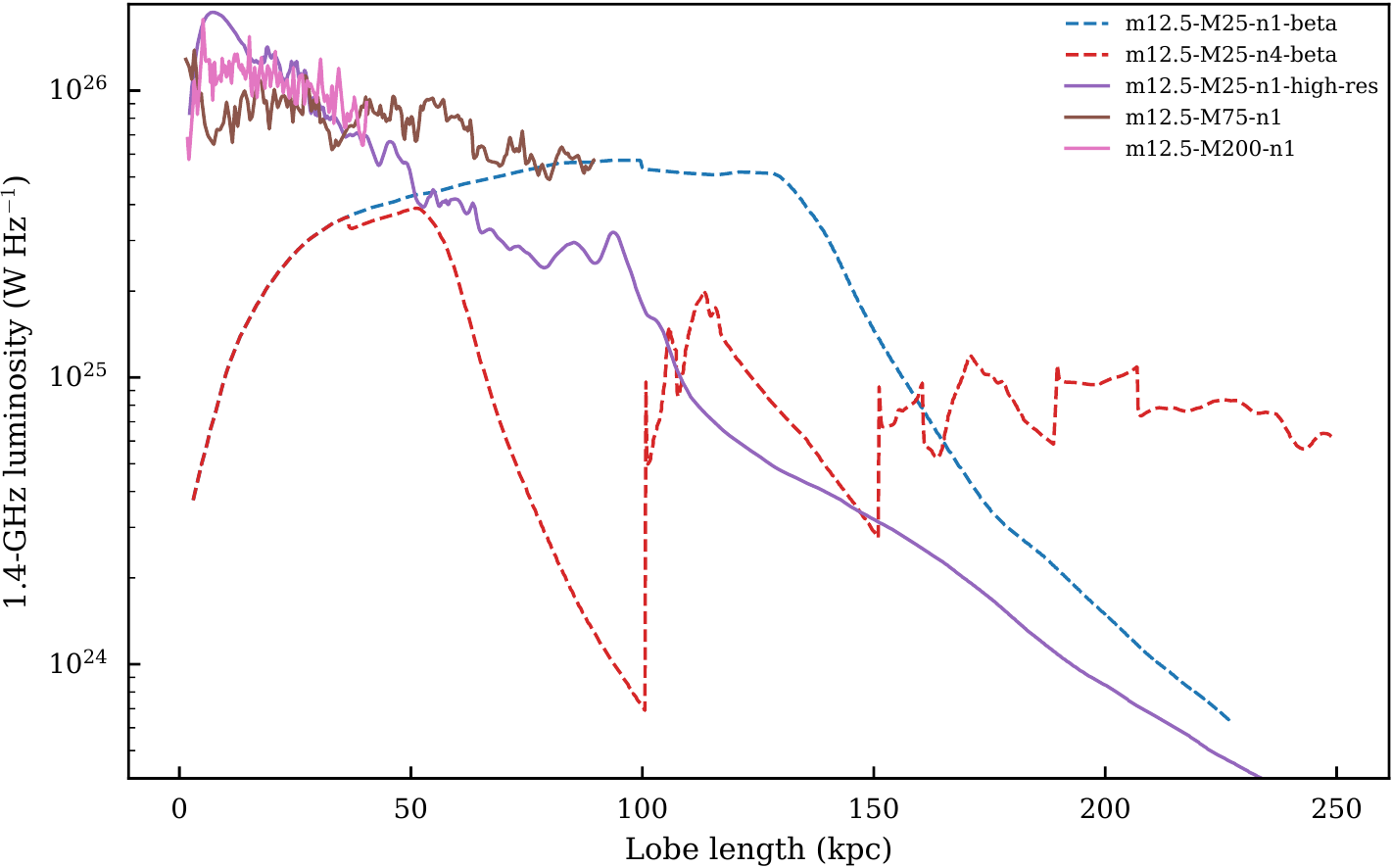}
    \caption{Size-luminosity diagram for supplementary (high resolution and high Mach number) simulations in the poor group environment.}
    \label{fig:pd-pg-extra}
\end{figure*}

\begin{figure*}
    \centering
    \begin{subfigure}[t]{0.5\textwidth}
        \centering
        \includegraphics{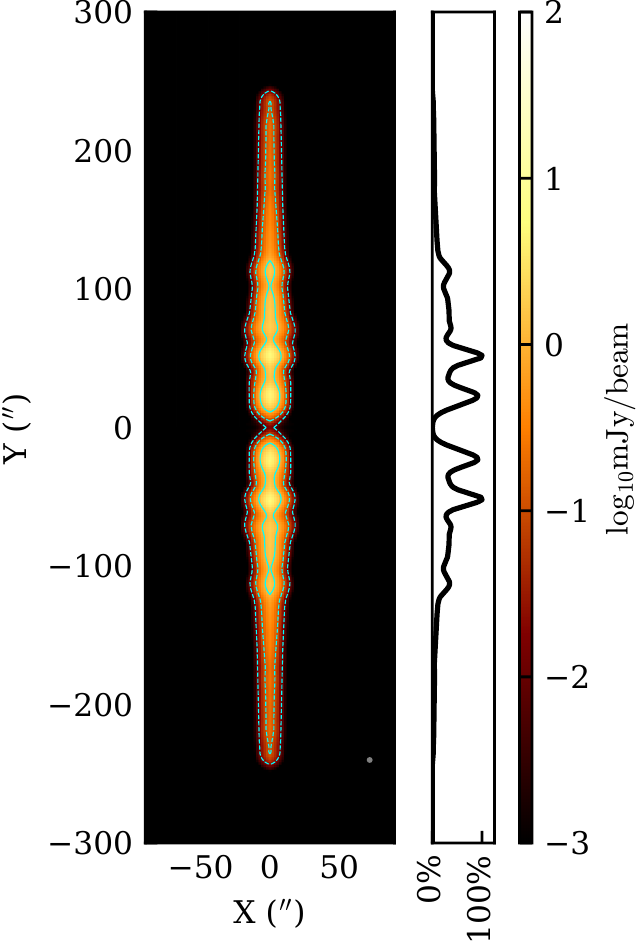}
        \caption{Cluster}
    \end{subfigure}%
    ~
    \begin{subfigure}[t]{0.5\textwidth}
        \centering
        \includegraphics{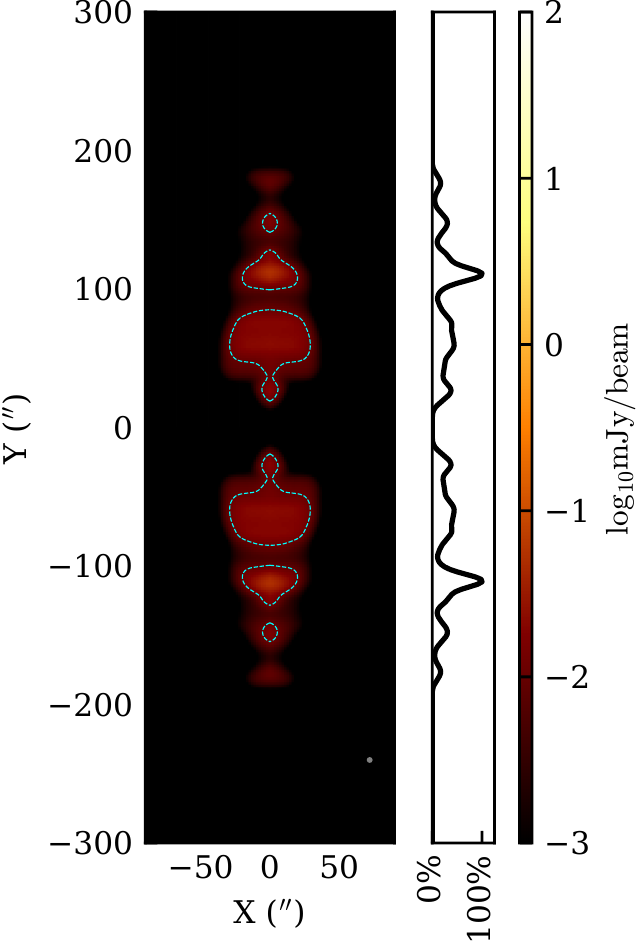}
        \caption{Poor group}
    \end{subfigure}%
    \caption{Surface brightness plots for 4 outburst jet at $t=200\,\mathrm{Myr}$. As in \autoref{fig:sb-n1-cluster-group}, radiative losses are neglected in the radio model, and so the surface brightness maps are upper limits. Contours are at $0.01$, $0.1$, $1$, $10$, $100$ $\mathrm{mJy}/\mathrm{beam}$}
    \label{fig:sb-n4-cluster-group}
\end{figure*}

As in \autoref{sec:env-pd} we plot the total luminosity against lobe length for the cluster and poor group environments, shown in \autoref{fig:pd-n4}.
The initial expansion period of the jets produces evolutionary tracks that match with their corresponding single outburst simulations, and then the total luminosity drops sharply when the jet is switched off.
For both the poor group and cluster environment, the peak luminosity of subsequent outbursts matches that of the first outburst within a factor of two.
Individual luminosity peaks are higher than the luminosity that would be expected for a jet that was active continuously.

As with the $n=1$ jets, there is a large evolution of the total luminosity over the simulation time, spanning approximately $2\,\mathrm{dex}$.
The number of outbursts that have previously occurred greatly affects the position of the source in the P-D diagram.
An active source may appear to have the total luminosity equal to or less than that of an inactive source, which has implications for the observations of double-double radio galaxies.

P-D tracks for all regular simulations in the cluster and poor group isothermal NFW environments are shown in \autoref{fig:pd-c} and \autoref{fig:pd-pg} respectively.
In all intermittent simulations the total luminosity is greater at larger lobe lengths, due to recent jet activity; similarly it is lower at shorter lobe lengths.
The finding that the peak luminosity of subsequent outbursts matches that of the first outburst within approximately a factor of two is confirmed for varying numbers of outbursts.
This indicates that the position of an observed source in the luminosity-size diagram will vary greatly depending on the intermittency of the underlying jet injection.

The surface brightness plots for the simulations with multiple outbursts are calculated as outlined in \autoref{sec:env-sb}.
\autoref{fig:sb-n4-cluster-group} shows surface brightness distributions for the 4-outburst simulations in the cluster and poor group environments, at $t=200\,\mathrm{Myr}$.
Multiple separate bubbles are present in the poor group environment, an indicator of intermittent jet activity; these should be visible in high sensitivity observations.
There are multiple peaks in surface brightness along the jet axis for the simulation in the cluster environment; however in an observation it would be difficult to confidently link these to intermittent jet activity, and would instead likely be attributed to knots in the jet.
This indicates that jet intermittency is likely one of several factors involved in producing double-double radio sources.
The multiple bow shocks discussed in \autoref{sec:int-jet-dynamics} would be visible in X-ray images as shown by \citet{Hardcastle2013}, however calculation of these is beyond the scope of this paper.

\begin{figure*}
    \includegraphics{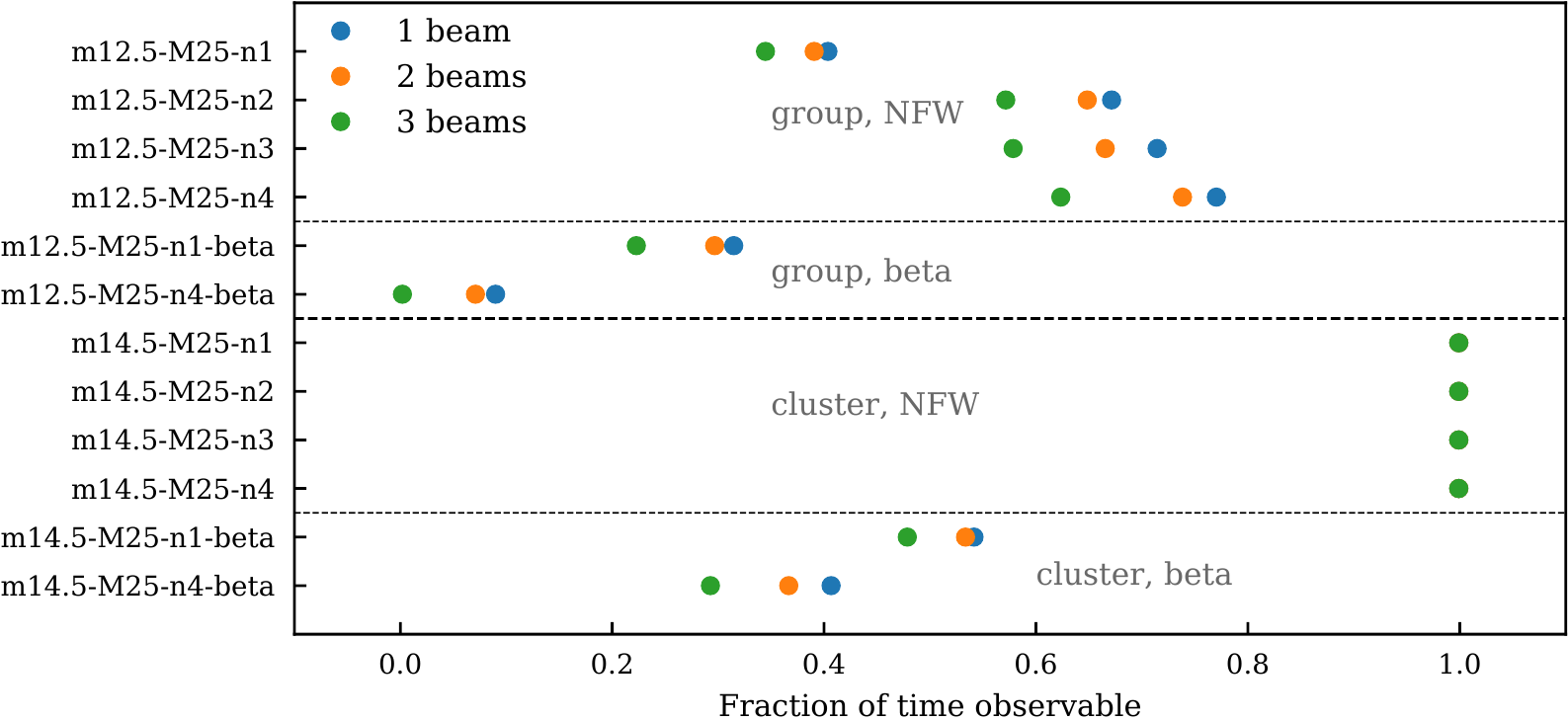}
    \caption{Fraction of time with at least 1, 2 or 3 beamwidths above $1\,\mathrm{mJy}/\mathrm{beam}$}
    \label{fig:observability-1mjy}
\end{figure*}

\begin{figure*}
    \includegraphics{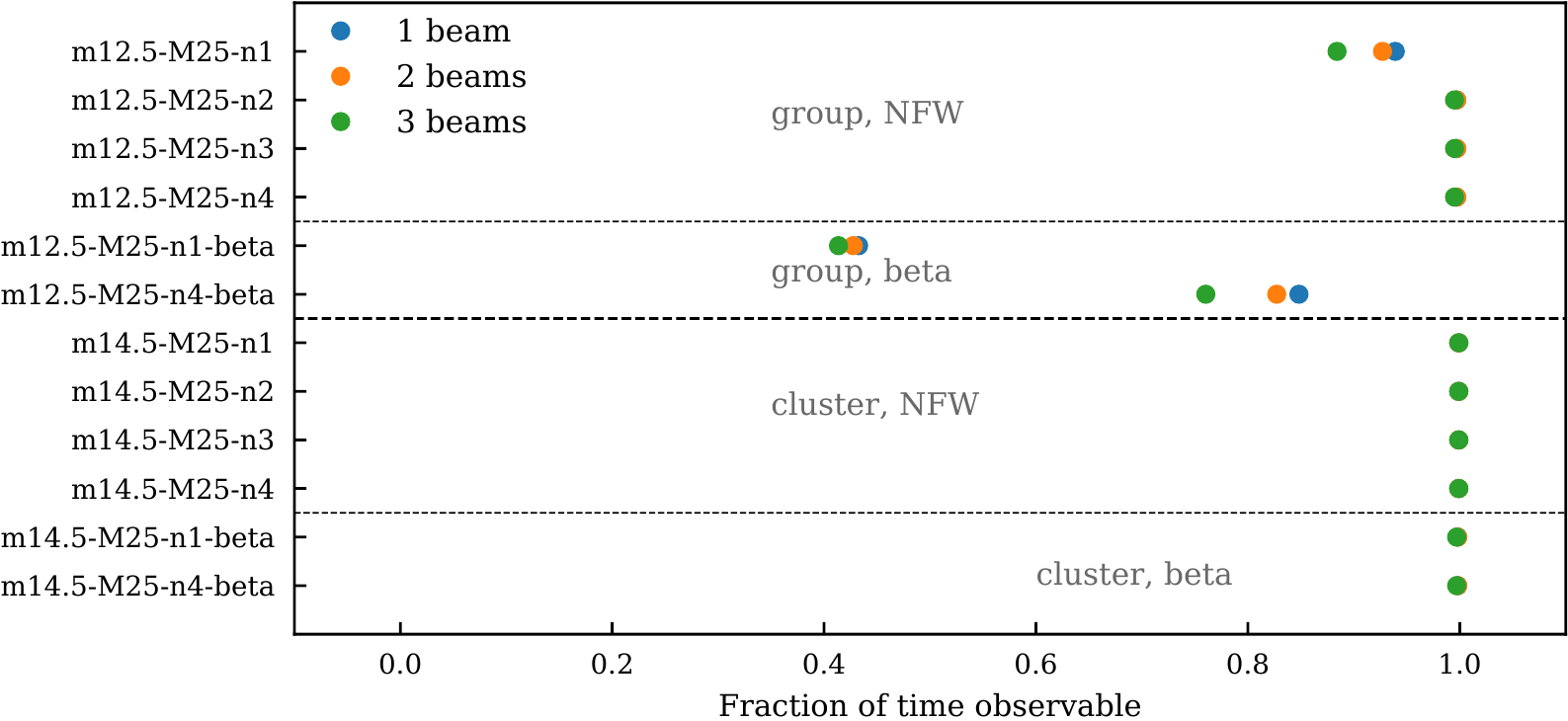}
    \caption{Fraction of time with at least 1, 2 or 3 beamwidths above $0.1\,\mathrm{mJy}/\mathrm{beam}$}
    \label{fig:observability-0.1mjy}
\end{figure*}

In order to quantify the effect intermittency has on the detectability of the radio source, we calculate the fraction of time it satisfies a detectability condition.
This detectability condition is defined to represent how confident an observer would be in associating the radio source lobes with the core emission (not calculated in our surface brightness maps).
With this in mind, the condition was chosen to be at least $x$ beamwidths across the jet above a certain surface brightness threshold $\alpha$.
The fraction of time spent satisfying the observability condition for each source is shown in \autoref{fig:observability-1mjy} and \autoref{fig:observability-0.1mjy}; here $x$ is 1, 2, and 3 respectively, corresponding to 1, 2, and 3 beams across the source, and we choose two surface brightness thresholds of $\alpha=1\,\mathrm{mJy}/\mathrm{beam}$ and $\alpha=0.1\,\mathrm{mJy}/\mathrm{beam}$ respectively.
The low sensitivity surface brightness threshold is chosen to roughly match that of the FIRST survey, while the high sensitivity surface brightness threshold is chosen to conservatively represent next generation radio surveys.

For the low sensitivity, $1$ beam condition, radio sources in the cluster isothermal NFW environment are detectable $100$ per cent of the time, compared to $77$ per cent of the time for sources in the poor group isothermal NFW environment.
A similar pattern is seen the $2$ and $3$ beam conditions.
Similarly for the high sensitivity conditions, radio sources in both isothermal NFW environments are nearly $100$ per cent detectable for all beamwidth conditions.
The simulated radio sources in the isothermal beta environments for both sensitivities are observable for a significantly smaller fraction of the simulation time, compared to their isothermal NFW counterparts.

These detectability fractions support the conclusion above that surveys sensitive to low surface brightness objects are important for discovering intermittent radio galaxies in poor environments.

\section{Discussion}\label{sec:discussion}

\begin{figure*}
    \includegraphics{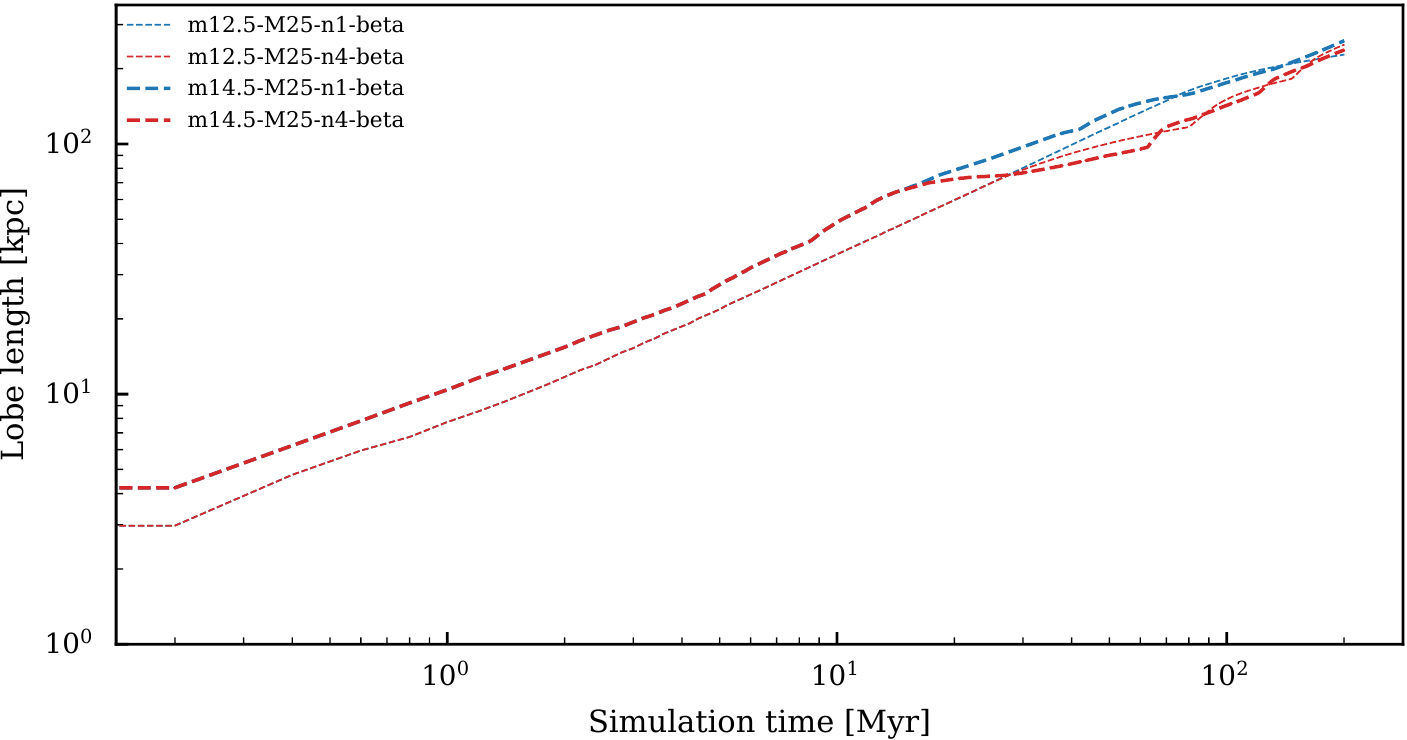}
    \caption{Evolution of lobe length for simulations in the cluster and poor group isothermal beta environments.}
    \label{fig:ll-all-beta}
\end{figure*}

\begin{figure*}
    \includegraphics{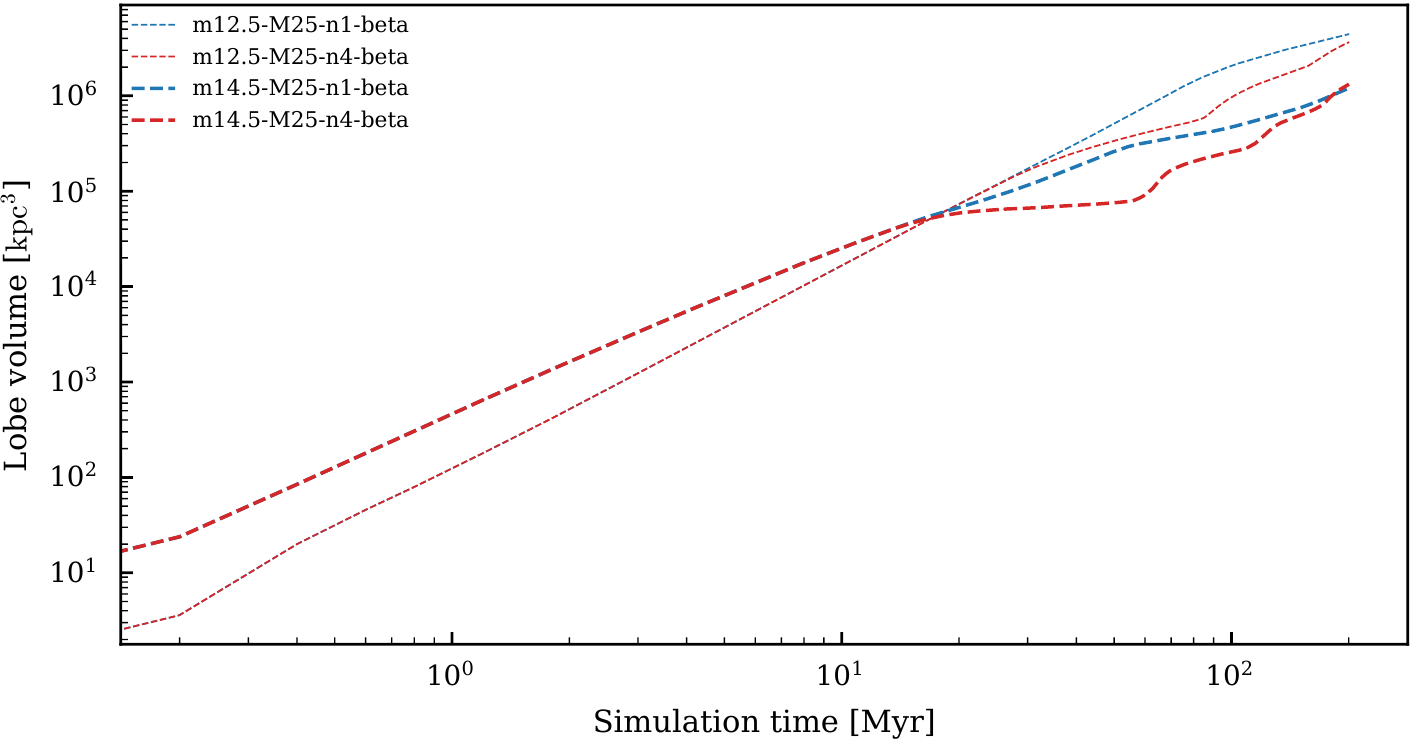}
    \caption{Evolution of lobe volume for simulations in the cluster and poor group isothermal beta environments.}
    \label{fig:lv-all-beta}
\end{figure*}

\begin{figure*}
    \includegraphics{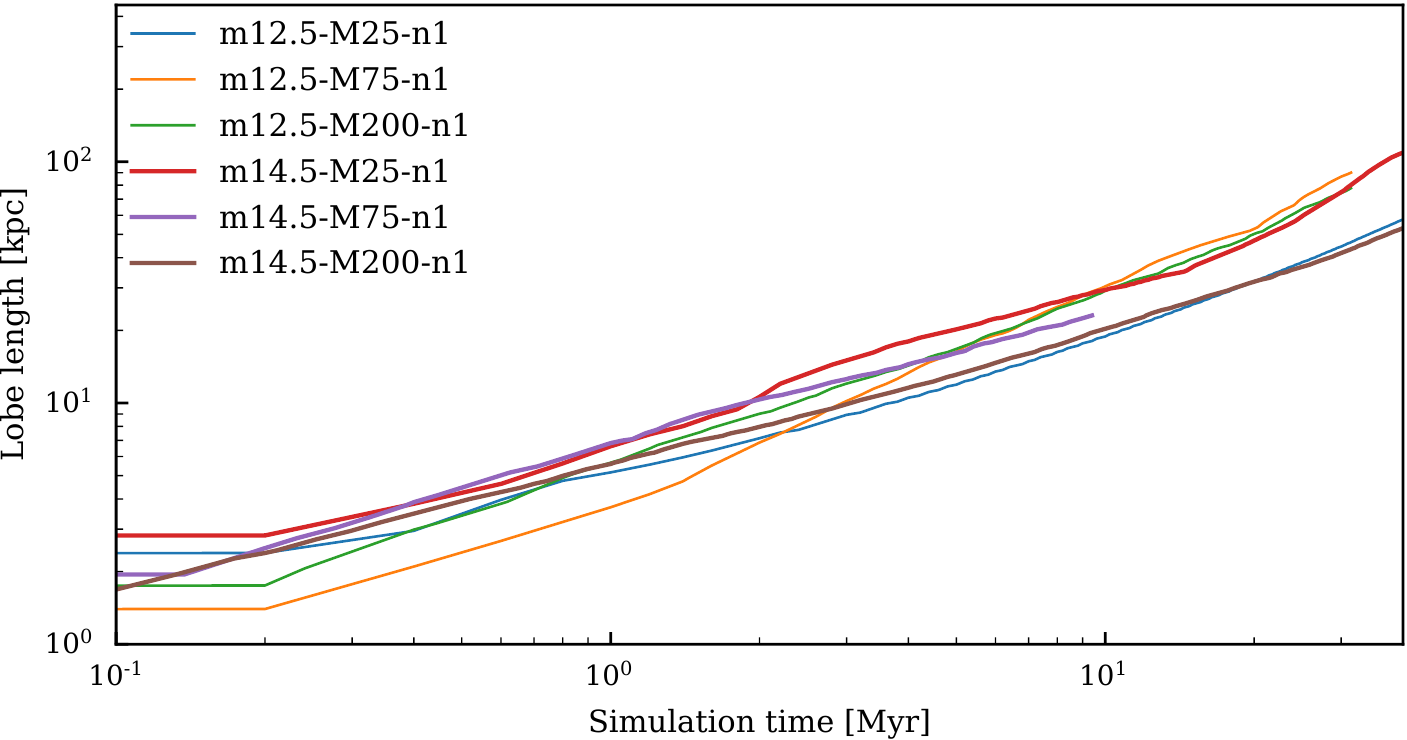}
    \caption{Length comparison for different Mach numbers}
    \label{fig:m-comp-length}
\end{figure*}

\begin{figure*}
    \includegraphics{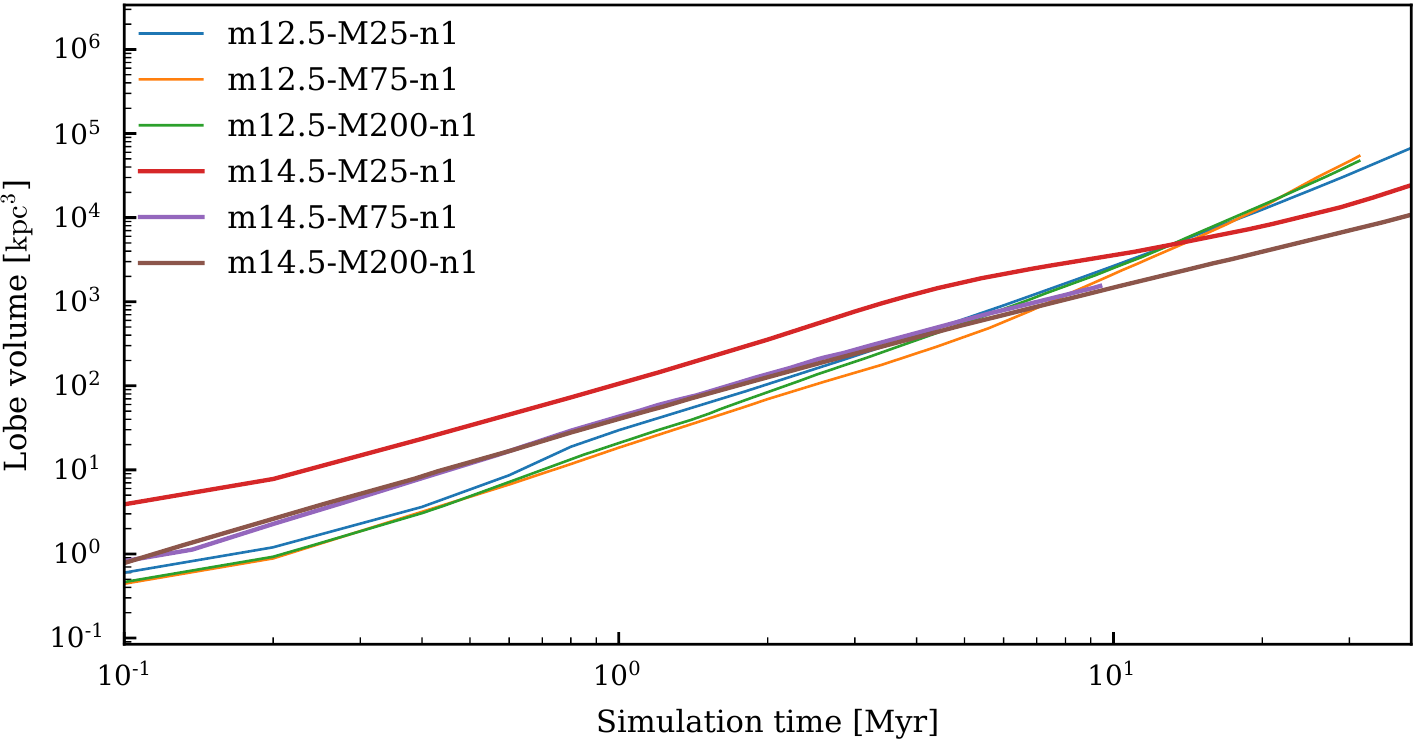}
    \caption{Volume comparison for different Mach numbers}
    \label{fig:m-comp-volume}
\end{figure*}

Several important technical aspects of the simulations need to be considered: whether the jet dynamics are accurately captured by the two-dimensionality of the simulations; whether the standard simulation resolution is sufficient to resolve the jet dynamics; and whether the chosen standard Mach number is sufficient for describing accurately the overall cocoon dynamics and radio observables.
This is done below.
Additionally we compare the standard isothermal NFW environment with the isothermal beta environment, and quantify the differences in jet dynamics and radio observables.
Next we consider the validity of our simulated synchrotron emission process, and then briefly discuss the reasoning behind our chosen standard jet kinetic power.
In summary, we believe our simulations and analysis have a systematic uncertainty of $60$ per cent in the length of a radio source, and a factor of up to $70$ per cent in luminosity.
For investigated dark matter halo masses, the jet power we use in our simulations should be an upper limit. Relative trends are more robust than absolute scalings, and conclusions in the previous sections remain qualitatively unchanged.

\subsection{Dimensionality}\label{sec:disc-dimensionality}

All simulations presented are two-dimensional axisymmetric (2.5D).
Two-dimensional simulations of FR II radio sources are able to reproduce the dynamics and overall lobe morphology of corresponding three-dimensional simulations, as shown by \citet{Hardcastle2013,Hardcastle2014}.
Also important for our simulations are entrainment rates of ambient gas into the lobe regions, because this re-filled gas is what the restarted jets work against.
Entrainment is related to the instabilities and turbulence, which are different in three dimensions.
The three dimensional nature of turbulence can even lead to jet disruption, especially if they have low power \citep{Massaglia2016}.
However, the dominant shear between radio lobes and shocked ambient gas in FR II sources simulated by this work is due to the backflow from the hotspot, and therefore, the entrainment should be reasonably well captured by our simulations, as previously shown by \citet{Hardcastle2014}.
The high-power FR II jets we discuss here are observed to be stable and this is typically also the case in our simulations.
A discussion of reasons for and complications with their stability is beyond the scope of this work.

The presence of turbulence in the environment is necessarily not axisymmetric, and so can only fully be explored with three dimensional simulations.
\citet{Bourne2017} investigated the effect of different levels of turbulence on jet morphologies and found that the jet inflated cocoons have a larger asymmetry when large-scale turbulence is present.

Our two-dimensional simulations also cannot capture the clumpy nature of the interstellar medium.
While important for quantifying feedback on galactic disk scales \citep{Mukherjee2016,Mukherjee2018preprint,Cielo2018}, the simulations presented here are appropriate for studies of large radio sources beyond the flood-and-channel phase. 

We also note again that our simulations do not include electron acceleration (e.g. at shocks) and loss (synchrotron and Inverse Compton cooling) processes.
Three-dimensional MHD simulations accounting for electron re-acceleration and loss processes \citep[e.g.][]{Jones1998} are needed for interpretation of finer-scale features, however our simulations are sufficient for a broad description of lobe dynamics. 

\subsection{Resolution}

The dependence of our results on resolution is checked by simulating higher resolution single outburst jets in the poor group and cluster isothermal NFW environments.
We can compare the dynamics of these simulations to the standard ones through lobe length and volume evolution, as shown in \autoref{fig:ll-all} and \autoref{fig:lv-all}.
We find that the high resolution simulation in the cluster isothermal NFW environment produces P-D tracks similar to the low resolution simulations, but lengths are consistently $\sim30$ per cent lower at a given age.
Meanwhile, the high resolution P-D tracks in the poor group have a marginally steeper slope.
Comparing lobe volume evolution, the high resolution simulations in both the poor group and cluster isothermal NFW environments have very similar slopes at later times (after $\sim10\,\mathrm{Myr}$), and differences of $\sim5$ per cent and $\sim25$ per cent at a given age to the corresponding standard simulation respectively.
This effect is well known from resolution studies in jet simulations \citep{Krause2001}: At higher resolution, the jet heads suffer more instabilities, spread out further, and hence the forward thrust is distributed over a larger working surface.
Even weak magnetic fields can stabilise the contact surface near the jet head \citep{Gaibler2009}.
Our lobe length calculations are hence systematically uncertain by about $30$ per cent.

The evolution of the radio source luminosity with size also provides a valuable tool for verifying our simulation resolution, as shown in \autoref{fig:pd-c-extra} and \autoref{fig:pd-pg-extra}.
Here our high resolution simulations are similar to the corresponding standard simulations for the same source size in both isothermal NFW environments.

\subsection{Mach number}

The final technical aspect of these simulations to discuss is the Mach number chosen for the jet.
As explained in \autoref{sec:jet-injection}, simulating the large-scale evolution of high Mach number jets on realistic time-scales is very computationally intensive.
Nevertheless two high Mach number simulations (Mach 75, Mach 200) were carried out for both the poor group and cluster isothermal NFW environments.
As the kinetic jet power $Q$ is constant, $\rho(r) v^3$ must be constant, and so $\rho \propto v^{-3}$.
This means that the ram pressure of the jet is $\rho v^2 \propto \frac{1}{v}$, and so higher Mach number jets are expected to collimate at smaller radii.
This smaller collimation radius is reproduced in the high Mach number simulations.
This means, while the higher Mach number jets (same jet power) have less forward ram pressure, they also distribute it over a smaller area.
Because of this, the dynamics of different Mach numbers are similar, and so we expect our analysis to be largely independent of the jet velocity.
We compare the dynamics of different Mach number simulations through lobe length and volume evolution, as shown in \autoref{fig:m-comp-length} and \autoref{fig:m-comp-volume} respectively.
We find that the lobe length and volume of our jets with different Mach numbers agree within $60$ per cent.

An additional verification for the independence of our analysis with respect to Mach number is the evolution of radio source luminosity with size (\hyperref[fig:pd-c-extra]{Figures \ref*{fig:pd-c-extra}} and \ref{fig:pd-pg-extra}).
There is good agreement between the Mach 75 and standard simulations for the cluster isothermal NFW environment, while the Mach 200 jet is significantly different by $\sim 70$ per cent, possibly due to the lobe volume being smaller for a given time (see \autoref{fig:m-comp-volume}).
The high Mach number simulations in the poor group isothermal NFW environment agree with the standard simulations up to a lobe length of $\sim50\,\mathrm{kpc}$, before starting to diverge by $\sim 20$ per cent due to the differences in lobe length evolution with time (see \autoref{fig:m-comp-length}).

\subsection{Environment profile}

The overall dynamics of radio sources simulated in the isothermal beta profile are similar to those using the isothermal NFW profile.
This is evident by comparing the length and volume of simulations in the isothermal beta environments (\hyperref[fig:ll-all-beta]{Figures \ref*{fig:ll-all-beta}} and \ref{fig:lv-all-beta}) with the corresponding isothermal NFW simulations (\hyperref[fig:ll-all]{Figures \ref*{fig:ll-all}} and \ref{fig:lv-all}).

Comparing the luminosity evolution of simulations using the isothermal NFW gas density to those using the isothermal beta profile (\hyperref[fig:pd-c]{Figures \ref*{fig:pd-c}}, \ref{fig:pd-c-extra}, \ref{fig:pd-pg} and \ref{fig:pd-pg-extra}) shows that while the total luminosity is $\sim1\,\mathrm{dex}$ lower for the cluster isothermal beta profile compared to the cluster isothermal NFW profile, the overall P-D track shape in the luminosity-size diagram is reproduced.
For the poor group simulations there is a larger difference between the two environment profiles ($\sim1-2\,\mathrm{dex}$ at larger sizes); this is due to the high central density and pressure in the poor group isothermal NFW environments.

\subsection{Synchrotron emission}

As described in \autoref{sec:env-pd} and \aref{app:synch-derivation}, the synchrotron emissivity per unit volume is calculated from the lobe pressure, assuming an equipartition factor for the electron and magnetic field energy density.
This method is similar to that used in \citet{Hardcastle2013}, and models the overall lobe luminosity reasonably well.
We do not explicitly model the synchrotron luminosity of the jet, however this is unlikely to affect our inferred luminosities and surface brightness distributions for two reasons: (i) jet luminosity typically constitutes only a small fraction of the overall luminosity of classical double radio sources \citep[e.g.][]{Hardcastle1998}; and (ii) once collimated, the jet will be pressure equilibrium with the cocoon plasma.
Because no radiative losses are included in our modelling, the synchrotron emissivity should be treated as an upper limit; this is a reasonable approximation for sources significantly younger than ~100 Myr \citep[e.g.][]{Kaiser1997a}, which is the case for all active phases in our simulations.

\subsection{Jet kinetic power}

The jet kinetic power studied in this paper ($Q=10^{37}\,\mathrm{W}$) was chosen to be in the FR I/II transition region, and plausible for jets in both poor groups and clusters.
A halo mass of $3\times10^{12}\,\mathrm{M}_\odot$ corresponds to a $\sim10^{11}\,\mathrm{M}_\odot$ galaxy \citep{Shabala2009,Croton2016a}.
Such a galaxy might reasonably have a $m_\mathrm{BH} = 10^{7}\,\mathrm{M}_\odot$ black hole \citep{Reines2015}, with an Eddington luminosity of $L_\mathrm{Edd}\sim10^{38}\,\mathrm{W}$.
We expect a maximum of $10$ per cent of the Eddington luminosity to be present in the jet \citep{Turner2015}.
Hence, we have simulated the most optimistic case for this kind of halo, and so our observability calculations are an upper bound. 

\section{Conclusions}\label{sec:conclusions}

We have shown that there is a clear link between the environment into which the jets are propagating, and the resulting morphology of the jet.
A large factor in this morphology difference is the collimation distance of the jet, which is larger in the poor group.
This results in a wider overall jet beam, and produces different large-scale structures.
Clearly, injecting the jet with a finite opening angle is an important factor for the radio morphology.
The environment affects the observable properties of the jets, as seen in the P-D tracks and surface brightness maps for jets in both the cluster and poor group environments.

Simulated radio observations of the jets show that the jet in the cluster is significantly easier to detect due to its higher surface brightness.
Comparing the two surface brightness distributions in \autoref{fig:sb-n4-cluster-group}, detecting emission from the radio lobes in the group environment would be difficult and possibly only the compact core would be visible, whereas the cluster environment has easily detectable extended emission.
The detectability of a simulation is quantified in \autoref{fig:observability-1mjy} and \autoref{fig:observability-0.1mjy}, where cluster radio sources are detectable up to $100$ per cent of the time (for a FIRST-like detection threshold) using our adopted parameters, while poor group environments are detectable at most $60$ per cent of the time.
Increasing the sensitivity by an order of magnitude allowed both cluster and poor group radio sources to be detectable up to $100$ per cent of the time.
This agrees with the findings presented by \citet{Shabala2008,Shabala2018} that massive galaxies (often residing in big haloes) host a larger fraction of extended radio sources.
It is expected that next generation radio surveys will detect a greater population of radio sources in poorer environments, due to increased sensitivity.
Future simulations for a range of jet powers and environments would allow the P-D diagram to be fully explored and could provide a framework to link observations to the underlying jet properties, aiding in placing radio sources of all sizes, including the ubiquitous compact sources \citep{Sadler2014a,Baldi2015,Shabala2017} on an evolutionary sequence.

The intermittency of a jet also plays a role in determining its large-scale morphology and observable properties.
Interestingly, the radio sources in subsequent active phases reach a similar radio luminosity to the first outburst, due to efficient entrainment.
Intermittency of radio activity is likely responsible for double-double radio sources.
Further modelling work, together with high-sensitivity, low-frequency observations \citep{Shimwell2017,Brienza2017} will shed light on the physics of this population.

Future work would include simulating a wide range of jet powers, environments, and opening angles, as well as FR I morphologies.
Our simulations presented in this paper only focused on producing radio sources with an FR II morphology.
Radio sources with an FR I morphology have a different (core-brightened) surface brightness profile, which has direct implications for observed source sizes and integrated luminosities.
Similar simulations of lower jet powers typical of FR I jets would provide information on how the observable properties of the FR I jets change due to jet-environment interaction and complement the results presented here.

\section*{Acknowledgements}

PMY thanks the University of Tasmania for a Dean's Summer Research Studentship and an Australian Postgraduate Award.
SSS and MGHK thank the Australian Research Council for an Early Career Fellowship (DE130101399).
We also thank an anonymous referee for their useful comments.
This research was carried out using the Tasmanian Partnership for Advanced Computing high-performance computing clusters.

%\afterpage{\clearpage}

%%%%%%%%%%%%%%%%%%%%%%%%%%%%%%%%%%%%%%%%%%%%%%%%%%


\bibliographystyle{mnras}
\bibliography{./main.bib} % if your bibtex file is called example.bib

%%%%%%%%%%%%%%%%%%%%%%%%%%%%%%%%%%%%%%%%%%%%%%%%%%

%%%%%%%%%%%%%%%%% APPENDICES %%%%%%%%%%%%%%%%%%%%%

\appendix

\section{Synchrotron emission derivation}\label{app:synch-derivation}

The synchrotron emissivity per unit volume $J(\omega)$ can be written as

\begin{equation}
    \label{eqn:emissivity-unit-volume}
    J(\omega) = A \frac{\sqrt{3\pi} e^3 B}{16 \pi^2 \epsilon_0 c m_e (q+1)} \kappa \left( \frac{\omega m_e^3 c^4}{3 e B} \right) ^ {- \frac{q-1}{2}}
\end{equation}

with

\begin{equation}
    \label{eqn:emissivity-A}
    A = \frac{\Gamma \left( \frac{q}{4} + \frac{19}{12} \right) \Gamma \left( \frac{q}{4} - \frac{1}{12} \right) \Gamma \left( \frac{q}{4} + \frac{5}{4} \right) }{ \Gamma \left( \frac{q}{4} + \frac{7}{4} \right) }
\end{equation}

as shown in \citet[Chapter 8]{Longair2011}.
Here we have assumed a power-law distribution of electron energies $N(E)=\kappa E^{-q}$ with exponent $q$ and normalisation $\kappa$ at an angular frequency $\omega$, which relates to the observing frequency $\nu = \frac{\omega}{2\pi}$.
Throughout the rest of the analysis, we take $q=2.2$ as in \citet{Hardcastle2013}, which gives a spectral index $\alpha = \frac{1-q}{2} = -0.6$ that is typical of radio lobes.

The relationship between cocoon pressure and energy densities is given in \citet{Kaiser1997a} as

\begin{equation}
    \label{eqn:cocoon-pressure-energy-densities}
    p=(\Gamma_\textrm{c} - 1)(u_\textrm{e} + u_\textrm{B} + u_\textrm{T})
\end{equation}

where $p$ is the pressure, $u_\textrm{e}$, $u_\textrm{B}$ and $u_\textrm{T}$ are the electron, magnetic field and thermal energy densities respectively, and $\Gamma_\textrm{c}$ is the adiabatic index, taken to be $\Gamma_\textrm{c} = 4/3$ for a relativistic plasma.

The normalisation $\kappa$ can be written as

\begin{equation}
    \label{eqn:emissivity-kappa}
    \kappa = \frac{u_\textrm{e}}{I} = \frac{u_\textrm{B} \eta}{I}
\end{equation}

using $\eta = u_\textrm{B} / u_\textrm{e} = B^2 / (u_\textrm{e} 2 \mu_0)$, the ratio between the energy densities of the magnetic field and electrons respectively, and $I$ is the integral of $E N(E)$

\begin{equation}
    \label{eqn:emissivity-I}
    I = \int E \times E^{-q} dE = (m_e c^2)^{2 - q} (\gamma_\textrm{max}^{2-q} - \gamma_\textrm{min}^{2-q}) / (2-q)
\end{equation}

\autoref{eqn:cocoon-pressure-energy-densities} can be rewritten to give $u_\textrm{e}$ in terms of the cocoon pressure and departure from equipartition as

\begin{equation}
    \label{eqn:emissivity-electron-energy-density}
    u_\textrm{e} = \frac{p}{(\Gamma_\textrm{c} - 1) (\eta + 1)}
\end{equation}

with the assumption that there is no thermal energy.

Using \autoref{eqn:emissivity-electron-energy-density} and $u_\textrm{B} = B^2 / (2 \mu_0)$ one can write

\begin{equation}
    \label{eqn:emissivity-b-field}
    \begin{split}
        B &= (2 \mu_0 \eta u_\textrm{e})^{1/2}\\
          &= \left( \frac{2 \mu_0}{\Gamma_\textrm{c} - 1} \left[ \frac{\eta}{1+\eta} \right] \right)^{1/2}
    \end{split}
\end{equation}

which leads to an expression for $\kappa$ in terms of the cocoon pressure

\begin{equation}
    \label{eqn:emissivity-normalisation}
    k = \frac{1}{\Gamma_\textrm{c} - 1} \left( \frac{1}{1+\eta} \right) \frac{1}{I} p
\end{equation}

Substituting \autoref{eqn:emissivity-b-field} and \autoref{eqn:emissivity-normalisation} with $\nu = \omega / 2 \pi$ into \autoref{eqn:emissivity-unit-volume} gives

\begin{equation}
    \label{eqn:emissivity-hertz-unit-volume}
    J(\nu) = K(q) \left( \frac{e^3}{\epsilon_0 c m_e} \right) \left( \frac{\nu m_e^3 c^4}{e} \right)^{- \frac{q-1}{2}} (2 \mu_0)^{\frac{q+1}{4}} \frac{1}{I} \left( \frac{ \eta^{\frac{q+1}{4}} }{ (1+\eta)^{\frac{q+5}{4} }} \right) p^{\frac{q+5}{4}}
\end{equation}

\begin{equation}
    \label{eqn:emissivity-K}
    \begin{aligned}
        K(q) &= \left(\frac{A}{(\Gamma_\textrm{c} - 1)^\frac{q+5}{4}} \right) \left( \frac{2\pi}{3} \right)^{-\frac{q-1}{2}} \left( \frac{\sqrt{3\pi}}{16 \pi^2 (q+1) } \right)\\
        &= \left[ \frac{\Gamma \left( \frac{q}{4} + \frac{19}{12} \right) \Gamma \left( \frac{q}{4} - \frac{1}{12} \right) \Gamma \left( \frac{q}{4} + \frac{5}{4} \right) }{ \Gamma \left( \frac{q}{4} + \frac{7}{4} \right) } \right]%
        \frac{1}{(\Gamma_\textrm{c} - 1)^\frac{q+5}{4}}\\
        & \qquad \times \left( \frac{2\pi}{3} \right)^{-\frac{q-1}{2}} \left( \frac{\sqrt{3\pi}}{16 \pi^2 (q+1) } \right)
    \end{aligned}
\end{equation}

The luminosity in each simulation cell is then given by $L(\nu) = 4 \pi J(\nu) V$, where V is the cell volume.
The final luminosity scaled to physical units ($\mathrm{W\ Hz}^{-1}$) is then

\begin{equation}
    L(\nu) = L_0 \left(\frac{\nu}{1\,\mathrm{GHz}} \right)^{ - \frac{q-1}{2}} \left( \frac{ p_0}{10^{-11}\,\mathrm{Pa}} \right)^{\frac{q+5}{4}} \left( \frac{L_1}{\mathrm{kpc}} \right)^3
\end{equation}

where $L_0$ is the coefficient for $L(\nu)$ scaled to $(L_1, p_0, \nu) = (1\,\mathrm{kpc}, 10^{-11}\,\mathrm{Pa}, 1\,\mathrm{GHz})$.

%%%%%%%%%%%%%%%%%%%%%%%%%%%%%%%%%%%%%%%%%%%%%%%%%%

% Don't change these lines
\bsp	% typesetting comment
\label{lastpage}
\end{document}